\documentclass{pasj01}
\draft 
\Received{$\langle$reception date$\rangle$}
\Accepted{$\langle$acception date$\rangle$}
\Published{$\langle$publication date$\rangle$}

\begin{document}

\title{High-mass star formation possibly triggered by cloud-cloud collision in the H{\sc ii} region RCW~34}
\author{Katsuhiro Hayashi$^{1}$, Hidetoshi Sano$^{1,2}$, Rei Enokiya$^{1}$, Kazufumi Torii$^{3}$, Yusuke Hattori$^{1}$, Mikito Kohno$^{1}$, Shinji Fujita$^{1}$, Atsushi Nishimura$^{1}$, Akio Ohama$^{1}$, Hiroaki Yamamoto$^{1}$, Kengo Tachihara$^{1}$, Yutaka Hasegawa$^{4}$, Kimihiro Kimura$^{4}$, Hideo Ogawa$^{4}$ and Yasuo Fukui$^{1,2}$}%
\altaffiltext{1}{Department of Physics, Nagoya University, Chikusa-ku, Nagoya, Aichi, 464-8601, Japan}
\altaffiltext{2}{Institute for Advanced Research, Nagoya University, Chikusa-ku, Nagoya, Aichi, 464-8601, Japan}
\altaffiltext{3}{Nobeyama Radio Observatory, 462-2 Nobeyama Minamimaki-mura, Minamisaku-gun, Nagano 384-1305, Japan}
\altaffiltext{4}{Department of Physical Science, Graduate School of Science, Osaka Prefecture University, 1-1 Gakuen-cho, Naka-ku, Sakai, Osaka 599-8531, Japan}
\email{khayashi@a.phys.nagoya-u.ac.jp}


\maketitle

\begin{abstract}
We report a possibility that the high-mass star located in the H{\sc ii} region RCW~34 was formed by a triggering induced by a collision of molecular clouds. 
Molecular gas distributions of the $^{12}$CO and $^{13}$CO~$J=$2-1, and $^{12}$CO $J=$3-2 lines in the direction of  RCW~34 were measured by using the NANTEN2 and ASTE telescopes.
We found two clouds with the velocity ranges of 0--10 km~s$^{-1}$ and 10--14 km~s$^{-1}$.
Whereas the former cloud as massive as $\sim$1.4 $\times$ 10$^{4}$ ${\it M}_{\odot}$ has a morphology similar to the ring-like structure observed in the infrared wavelengths, the latter cloud with the mass of $\sim$600 ${\it M}_{\odot}$, which has not been recognized by previous observations, distributes just likely to cover the bubble enclosed by the other cloud.
The high-mass star with the spectral type of O8.5V is located near the boundary of the two clouds.
The line intensity ratio of $^{12}$CO $J=$3-2 / $J=$2-1 yields high values ($\gtrsim$ 1.0), 
suggesting that these clouds are associated with the massive star.
We also confirmed that the obtained position-velocity diagram shows a similar distribution to that derived by a numerical simulation of the supersonic collision of two clouds.
Using the relative velocity between the two clouds ($\sim$5 km~s$^{-1}$), the collisional time scale is estimated to be $\sim$0.2~Myr with the assumption of the distance of 2.5 kpc.
These results suggest that the high-mass star in RCW~34 was formed rapidly within a time scale of $\sim$0.2 Myr via a triggering of cloud-cloud collision.

\end{abstract}

\section{Introduction}

\subsection{High-mass star formation}

High-mass stars strongly affect environment of the interstellar medium (ISM) by injecting a large amount of energy via ultraviolet (UV) photons, stellar winds, and supernova explosions.
Materials around the massive stars are ionized by the UV radiation and are swept up by the stellar winds or supernova explosions, leading a formation of the H{\sc ii} region.
Heavy elements produced in the massive stars also have a great influence on the chemical evolution of the ISM. 
Investigating the formation process of massive stars and the environment around them are essential to elucidate the structure and evolution of galaxies. 
However, their remarkably short lives and the small number of population make observational constrains difficult.
Theoretical studies have remained controversial due to the lacking of observational guidance. 

In generally, ``core accretion'' and ``competitive accretion'' are thought be promising pictures to explain the formation process of high-mass stars (e.g., \cite{ZinneckerYorke07}; \cite{Tan+14}).
Both the two models, however, have not been established due to the lack of convincing confrontation between the theories and observations.
\citet{ElmegreenLada77} proposed a model that the massive stars are formed by the ambient material accumulated by  expanding motions of the ionized gas: 
high pressure in a shock wave driven by the expansion accumulates the surrounding material and forms a dense molecular layer, which becomes gravitationally unstable and collapses to form the next-generation of stars  (``collect \& collapse").
There are several candidates of high-mass star-forming regions suggesting the scenario of collect \& collapse (e.g., \cite{Deharveng+08}; \cite{Zavagno+10}). 

On the other hand, recent observational studies of the molecular clouds distributed in the star-forming regions suggest a possibility that high-mass stars would be formed by triggers of supersonic collisions between the interstellar clouds.
A large velocity difference ($\sim$10--30 km~s$^{-1}$) between these clouds is too large to be gravitationally bound, but they are likely physically associated as verified by the morphological correspondence and bridging features connecting two clouds in velocity space.
The excited gas distribution around the massive stars suggests physical associations between the clouds and massive stars. 
Molecular clouds having these properties are found in several massive star clusters, Westerlund~2, NGC~3603, RCW~38 and Orion Nebula Cluster (\cite{Furukawa+09}; \cite{Ohama+10}; \cite{Fukui+14}; \yearcite{Fukui+16}; \yearcite{Fukui+17}), as well as the H{\sc ii} regions which harbors a single O star, RCW~120 and M20 (\cite{Torii+15}; \yearcite{Torii+17}).
ALMA observations also found similar diagnostics in the high-mass star-forming regions in the Large Magellanic Cloud, N159 West, N159 East, and R~136 (\cite{Fukui+15}; \cite{Saigo+17}).
The typical time scale of the collision is estimated as $\sim$0.1--1~Myr, suggesting that the high-mass stars are formed rapidly within a few 10$^{5}$~years.  

Theoretical studies also showed that the cloud collision can be a trigger of the formation of massive stars (\cite{HabeOhta92}; \cite{Anathpindika10}; \cite{Takahira+14}; \cite{Haworth+15a}; \yearcite{Haworth+15b}).
Magnetohydrodynamical simulations derived that the collision induces formation of dense self-gravitating clumps at the compressed region, leading to a large mass accretion rate (10$^{-4}$ to 10$^{-3}$ ${\it M}_{\odot}$~yr$^{-1}$) enough to create massive stars \citep{InoueFukui13}.   
Numerical simulations also showed that the collision between the different size of clouds creates a cavity in the larger cloud, forming a ring-like gas distribution (e.g., Figure 12 in  \cite{Torii+17}).
After forming the massive stars, the surrounding materials are heated by the UV radiation, 
which are observed well in the infrared wavelengths (e.g., \cite{Torii+15}).
The cloud-cloud collision is an alternative view to give an unified explanation to form massive stars and the surrounding structures such as the infrared ring. 

\subsection{RCW 34 in the Vela Molecular Ridge}

\begin{figure}[h]
 \begin{center}
 \centering
  \includegraphics[width=10.0cm]{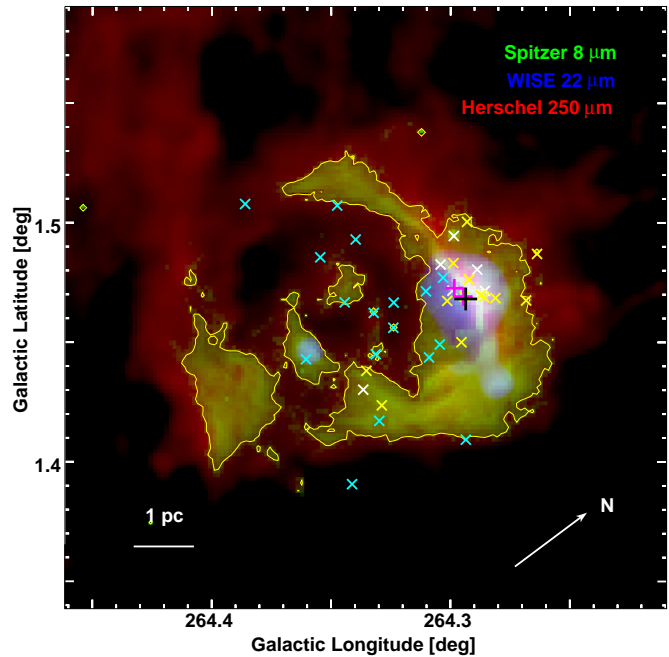}
  \end{center}
 \caption{Color composite image of the infrared emission from RCW~34. Green, blue and red indicate {\it Spitzer}/IRAC 8 $\micron$, {\it WISE} 22~$\micron$ and {\it Herschel}/SPIRE 250 $\micron$ distributions, respectively. Black cross shows the position of O8.5V exciting star and magenta ones are the positions of early B type stars. Yellow, white and cyan crosses indicate the positions of class 0/I, class I/II and class II objects, respectively \citep{Bik+10}. Yellow outlines in the 8 $\micron$ ring is used to explain to the 8 $\micron$ distribution in Figure~\ref{fig:12CO21and13CO21}.}
\label{fig:ThreeCompositeColorMap}  
\end{figure}

RCW~34 is one of the H{\sc ii} regions located in the Vela Molecular Ridge, a famous star forming region on the southern galactic plane, consisting of several giant molecular clouds (e.g., \cite{Yamaguchi+99}). 
It is characterized by a ring-like structure extended in a size of 4~pc $\times$ 7~pc at a distance of 2.5~kpc \citep{Bik+10}, accompanying a single O star, vdBH~25a (O8.5V, \cite{Hydari-Malayeri88}; \cite{Bik+10}) and two early B type stars (B0.5V and B3V, \cite{Bik+10}) located near the peak of the infrared emission.   
Figure~\ref{fig:ThreeCompositeColorMap} shows a composite color image of RCW~34, in which the positions of O / early B stars and the young stellar objects (YSOs) classified into class~0/I, class~I/II and class~II \citep{Bik+10} are overlaid.
One can see a large bubble enclosed by the infrared ring observed by {\it Spitzer}/IRAC 8 $\micron$ (green) and {\it Herschel}/SPIRE 250 $\micron$ (red), which mainly traces the polycyclic aromatic hydrocarbon (PAH) and cold dust, respectively.
The {\it WISE} 22 $\micron$ (blue) emission probably tracing hot dust are extended around the massive stars. 
While the massive stars and the class~0/I and I/II objects are located within the 8 $\micron$ ring, the class~II objects are distributed extensibly in the bubble.

\citet{Hydari-Malayeri88} conducted a detailed optical study, identifying the spectral type of the O star, and suggested high velocity flow (champagne flow) from the exciting star toward the southern direction by a measurement of H$\alpha$. 
As a complementary study, CO observations of the molecular clouds toward RCW~34 were carried out by \citet{Pagani+93}. 
They revealed gas distribution partly in front of the ionized gas, supporting the large visual extinction at the northern part of the bubble.
By using radio and near-infrared data obtained with {\it VLT} spectroscopic and {\it Spitzer} photometric observations, \citet{Bik+10} found that RCW~34 consists of three different regions according to distributions of the YSOs.
\citet{vanderWalt+12} revealed more detailed stellar distribution across the entire H{\sc ii} region including fainter low-mass stars traced by the near infrared observation with IRSF (InfraRed Survey Facility) telescope.

\citet{Bik+10} proposed a scenario for the formation of RCW~34:
a sequential star formation in the bubble from the south to north triggers the formation of the massive stars, which promotes current star formations in the dense molecular clouds.
However, the class II objects in the bubble show spatially extended distribution rather than clustering one zone (see Figure~\ref{fig:ThreeCompositeColorMap}).
It is difficult to find the age gradient from these stellar objects with the current data accuracy.   
As suggested by \citet{vanderWalt+12}, other star-forming scenario should be considered, if the distribution of the low-mass YSOs are found to be spread out in almost uniformly across the H{\sc ii} region.
Revealing the initial condition to trigger the sequential star formation is also a remained issue.

In the Vela Molecular Ridge, the youngest super star cluster RCW~38 is suited at $\sim$5$^{\circ}$ away from   RCW~34.
Recent observational studies of molecular clouds found that the formation of massive stars in RCW~38 can be explained by a triggering induced by cloud-cloud collision \citep{Fukui+16}.
The Vela Molecular Ridge has several H{\sc ii} regions showing the collisional diagnostics which may form the massive stars (\cite{Sano+17}, and Enokiya et al. 2017 in prep).
RCW~34, characterized by the ring-like structure with a single O star (and two early B type stars) might be also explained by a cloud collision, as suggested by other star forming regions, RCW~120 and M20 having the similar number of massive stars and cloud mass (\cite{Torii+15}; \yearcite{Torii+17}).  
In addition, detailed studies of molecular clouds in RCW~34 using recent millimeter/sub-millimeter observatories have not been reported yet.  
With the molecular transition lines of $^{12}$CO and $^{13}$CO $J=$2-1 data obtained by NANTEN2, and $^{12}$CO $J=$3-2 data provided by the ASTE observation, we investigate a possibility of the formation of high-mass star and the surrounding  structure via the cloud-cloud collision.

This paper is organized as follows. In Section \ref{sec:observation}, we describe the observations of the molecular clouds in RCW~34 by the NANTEN2 and ASTE telescopes. 
Section~\ref{sec:results} shows results obtained from the $^{12}$CO, $^{13}$CO $J=$2-1 and $^{12}$CO $J=$3-2 data.
In Section~\ref{sec:discussion}, we discuss a possibility of the cloud-cloud collision by a comparison with a simulation result and studies of the other star-forming regions.    
We give a conclusion of our study in Section~\ref{sec:conclusion}.
In this paper, the minimum contour level is set to be 5 $\sigma$ unless otherwise noted.  

\clearpage

\section{Observation}
\label{sec:observation}

\subsection{NANTEN2 $^{12}$CO and $^{13}$CO $J=$2-1 Observations}
\label{sec:nanten2coj2to1}
With the NANTEN2 4~m millimeter/sub-millimeter telescope located in Chile, a simultaneous $^{12}$CO and $^{13}$CO $J=$ 2-1 observation toward RCW~34 was conducted from October 31 to November 2 in 2015. 
The OTF (on-the-fly) mode was used for the mapping observation with orthogonal directions to reduce the scanning effect, which provides the cubic data covering the scale of \timeform{15'} $\times$ \timeform{15'}.
The system temperature including atmosphere was $\sim$130~K in the double-side band.  
A Fourier digital spectrometer installed on the backend of the beam transmission system provided data resolved into 16384 channels at 1GHz bandwidth.
The velocity range covers 650~km~s$^{-1}$ at 230~GHz, giving the frequency resolution of 61~kHz, which corresponds to the velocity resolution of 0.08~km~s$^{-1}$.
The pointing accuracy was measured to be $\lesssim$ \timeform{10''} by observing IRC+10216 located at (R.A., Dec.)=(\timeform{09h13m57.40s}, \timeform{+13D16'04.5''}).
The absolute intensity was estimated by a calibration using the CO data from Orion-KL at (R.A., Dec.)=(\timeform{05h35m14.48s}, \timeform{-5D22'27.55''}), which were obtained by the large-scale survey of the Orion molecular clouds performed by \citet{Nishimura+15}.
These data were smoothed with a Gaussian of $\sigma =$ \timeform{45''}, which gives the beam size of \timeform{90''} on the spatial map, and were smoothed to be 0.5~km~s$^{-1}$ in the velocity axis. 
The typical rms noise levels for the $^{12}$CO and $^{13}$CO $J=$ 2-1 data at the smoothed velocity resolution of 0.5~km~s$^{-1}$ were $\sim$0.3~K per channel and $\sim$0.5~K per channel, respectively.

\subsection{ASTE $^{12}$CO $J=$3-2 Observation}
\label{sec:astecoj2to1}
The observation of $^{12}$CO $J=$ 3-2 rotational transition line toward RCW~34 were carried out in November 2015, by using the ASTE 10~m sub-millimeter telescope situated in Chile (\cite{Ezawa+04}, \yearcite{Ezawa+08}).
The OTF mapping observation covered the \timeform{7.5'}~$\times$~\timeform{7.5'} filed of view arranged almost centered on the exciting star, vdBH~25a.
The typical system temperature including the atmosphere was measured to be $\sim$350--550~K in the single-side band.
The digital spectrometer ``MAC''\footnote{https://alma.mtk.nao.ac.jp/aste/instruments/mac.html} with the total bandwidth of 128~MHz and 0.125 MHz resolution at 350 GHz provided the data covering the velocity range of 111~km~s$^{-1}$ with the velocity resolution of 0.11~km~s$^{-1}$.
The pointing accuracy was checked by the observing RAFGL 4078 located at (R.A., Dec.)=(\timeform{07h45m02.41s}, \timeform{-71D19'45.7''}) every $\sim$1 hour to achieve an offset better than \timeform{3''}.
The absolute intensity was calibrated by using the measurements of the CO line from a standard source IRC+10216\footnote{http://alma.mtk.nao.ac.jp/aste/guide/prepare/callist\_en.html}.
Finally, these dada were smoothed to be 0.3~km~s$^{-1}$ in the velocity axis and to be \timeform{24''} on the spatial map.
The typical rms noise level was $\sim$0.3~K per channel at the smoothed velocity resolution of 0.3~km~s$^{-1}$.

\clearpage

\section{Results}
\label{sec:results}

\subsection{Two Clouds at Different Velocities}
\label{sec:RedandBlueClouds}

\begin{figure}[h]
 \begin{tabular}{cc}
  \begin{minipage}{0.5\hsize}
   \begin{center}
    \rotatebox{0}{\resizebox{8cm}{!}{\includegraphics{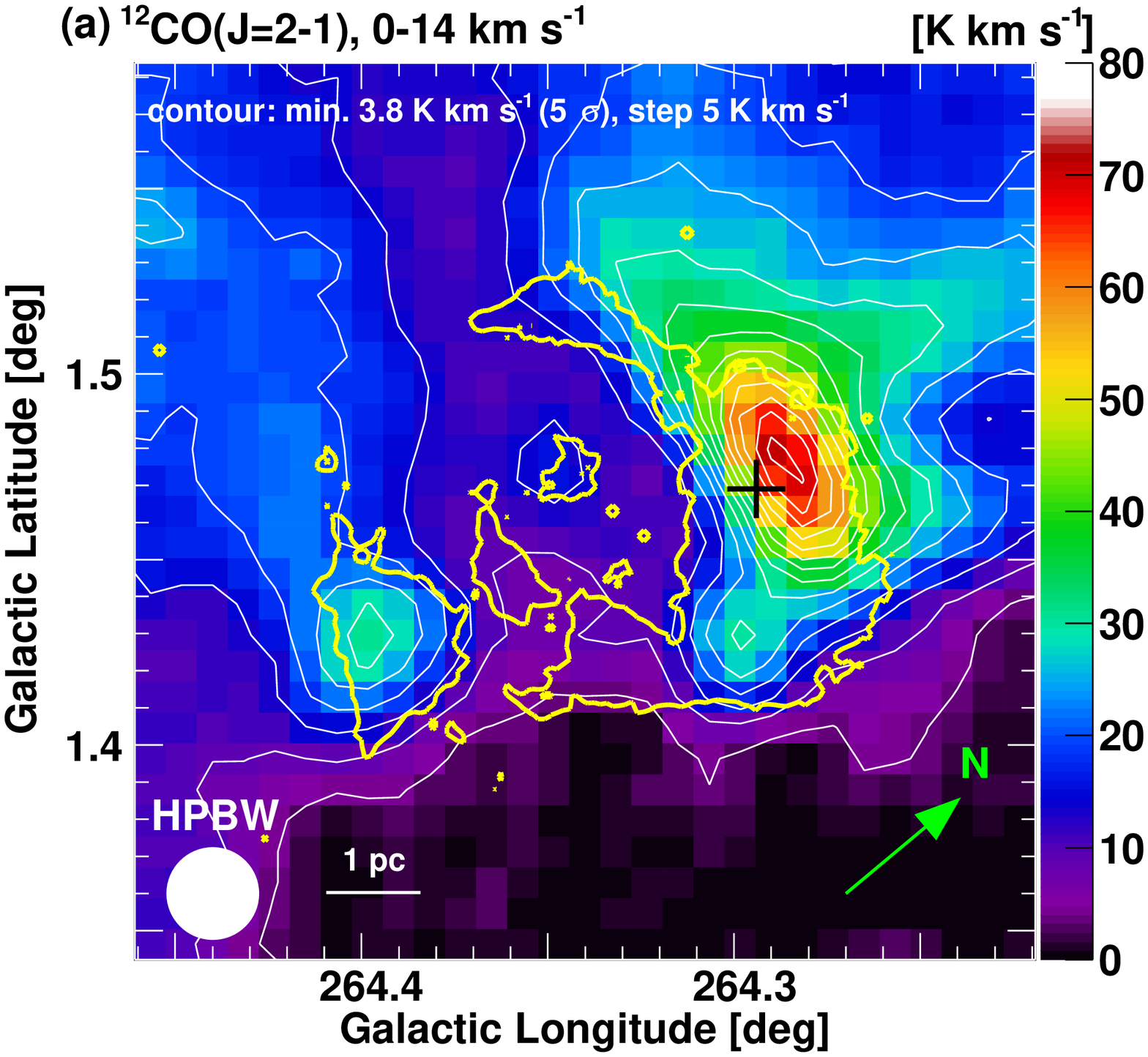}}}
   \end{center}
  \end{minipage} 
  \begin{minipage}{0.5\hsize}
   \begin{center}
    \rotatebox{0}{\resizebox{8cm}{!}{\includegraphics{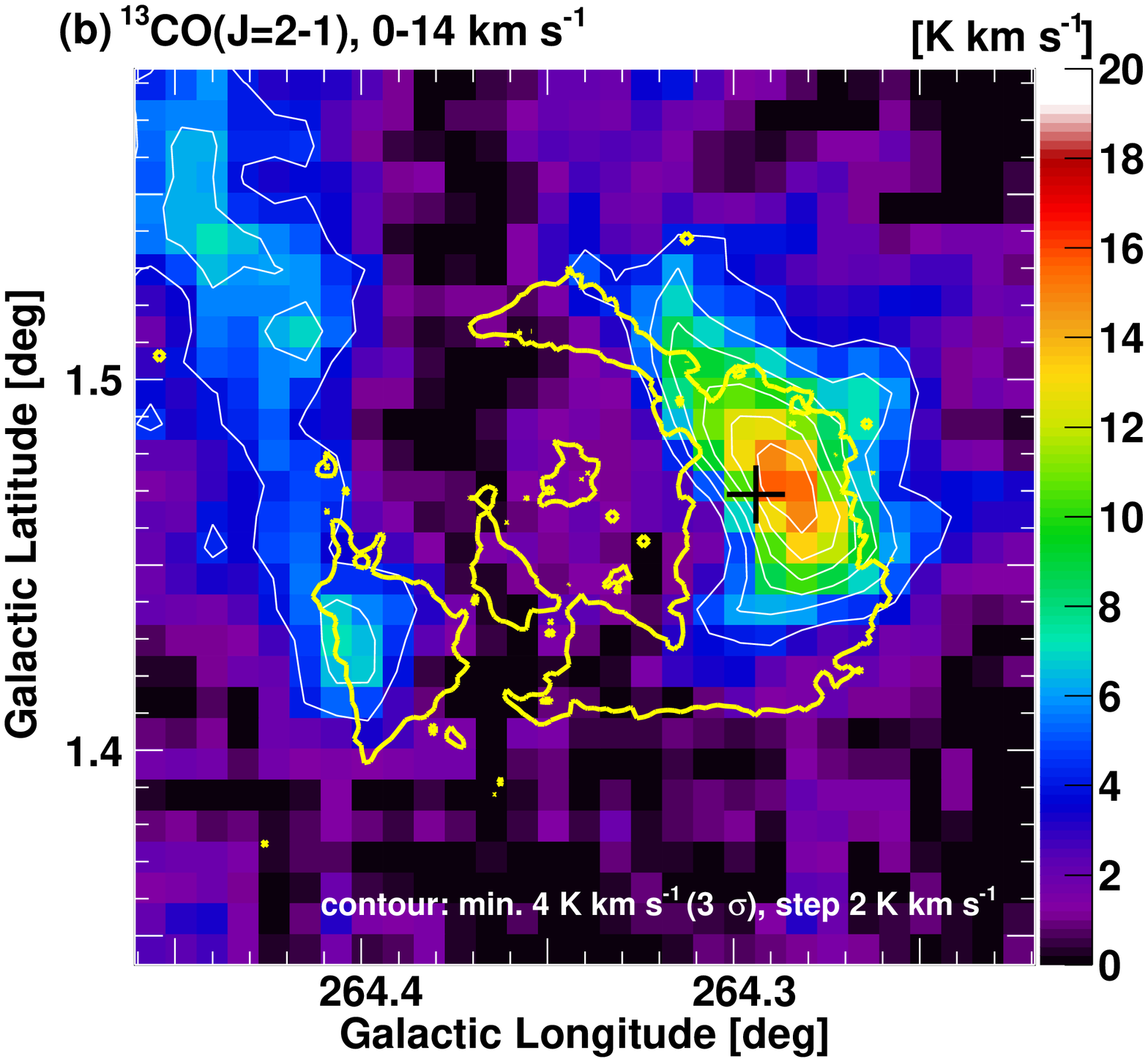}}}
   \end{center}
  \end{minipage} \\
  \end{tabular}  
  \caption{Integrated intensity distributions of the (a) $^{12}$CO $J=$2-1 and (b) $^{13}$CO $J=$2-1 lines with the velocity range from 0 to 14~km~s$^{-1}$. The cross indicates position of the O8.5V star. The yellow line indicates an outline of the 8 $\micron$ emission shown in Figure~\ref{fig:ThreeCompositeColorMap}.}
\label{fig:12CO21and13CO21}   
\end{figure}

\begin{figure}[h]
 \begin{center}
 \centering
  \includegraphics[width=16.0cm]{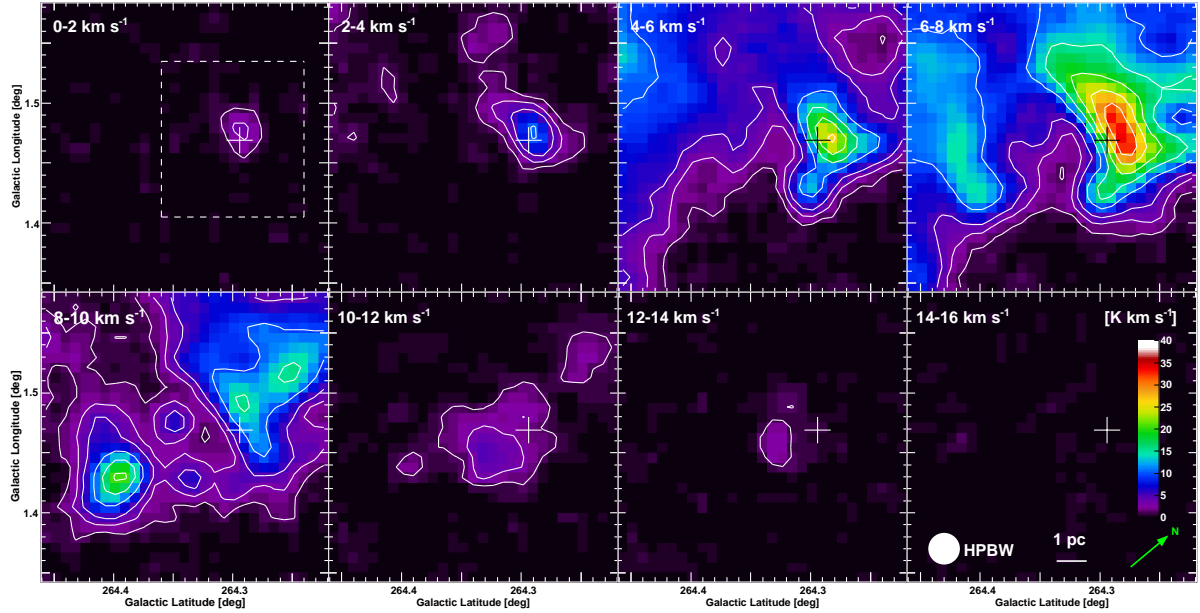}
  \end{center}
 \caption{Velocity channel distributions of the $^{12}$CO $J=$2-1 data. The contour levels are set to be 1.4, 3, 5, and 10~K~km~s$^{-1}$, and followed by increasing 5~K~km~s$^{-1}$ step. The cross indicates the position of O8.5V star. The white dashed box shown in the top-left panel indicates the analyzed region of the $^{12}$CO $J=$3-2 data obtained by ASTE (see Figure~\ref{fig:12CO32ChannelMap}).}
\label{fig:12CO21ChannelMap}  
\end{figure}

Figures~\ref{fig:12CO21and13CO21}(a) and (b) indicate the intensity distributions of $^{12}$CO and $^{13}$CO $J=$2-1 lines, respectively, with the integrated velocity ranges are 0 to 14~km~s$^{-1}$.
We found similar gas distributions to the infrared 8~$\micron$ ring shown by the yellow line.
The cloud in the northern area shows an arc-shaped morphology elongated to the east-west direction, which contains the main peak at ($l$, $b$) $\sim$ (\timeform{264D.29}, \timeform{1D.48}). 
Another peak is detected at ($l$, $b$) $\sim$ (\timeform{264D.4}, \timeform{1D.43}) in a cometary-like cloud distributed in the southern area. 
The bubble-like structure enclosed by the infrared ring (see Figure~\ref{fig:ThreeCompositeColorMap}) is also confirmed in the CO distribution.
The exciting star, vdBH~25a, with the spectral type O8.5V, is located at the slightly south from the highest CO peak.
The CO intensity around the O star is drastically changed as represented by the white contours with high contrast. 
From the $^{12}$CO data, we found low-intensity diffuse gas enclosed by the infrared ring, whereas the $^{13}$CO data did not show significant intensities (even if 3~$\sigma$ detection level).
To see the gas distribution in narrower velocity ranges, we made a velocity channel map separated into every 2~km~s$^{-1}$ with ${\it V}_{\rm LSR}$ (velocity of the local standard of reset) 
from 0~km~s$^{-1}$ to 14~km~s$^{-1}$ for the $^{12}$CO data as shown in Figure~\ref{fig:12CO21ChannelMap}.
We found that the gas distribution is mainly separated into two components with the different velocities, ${\it V}_{\rm LSR}$~$=$ 0--10~km~s$^{-1}$ and 10--14~km~s$^{-1}$ (hereafter denoted by ``blue cloud" and ``red cloud", respectively). 
The blue cloud exhibits the ring-like gas distribution which includes  
the highest intensity peak at ${\it V}_{\rm LSR}$~$\sim$7~km~s$^{-1}$, and low signifiant intensity gas extended in the bubble, such as the diffuse component at ${\it V}_{\rm LSR}$~$\sim$7~km~s$^{-1}$ and small clumpy structures with the size of $\sim$1 pc at ($l$, $b$) $\sim$ (\timeform{264D.35}, \timeform{1D.475}) and (\timeform{264D.34}, \timeform{1D.42}) at ${\it V}_{\rm LSR}$~$\sim$9~km~s$^{-1}$. 
The red cloud with the low intensity emission distributes just likely to cover the northern part of the bubble, which has not been recognized by the previous molecular observations probably due to low detection sensitivity.

\begin{figure}[h]
 \begin{tabular}{cc}
  \begin{minipage}{0.5\hsize}
   \begin{center}
    \rotatebox{0}{\resizebox{8cm}{!}{\includegraphics{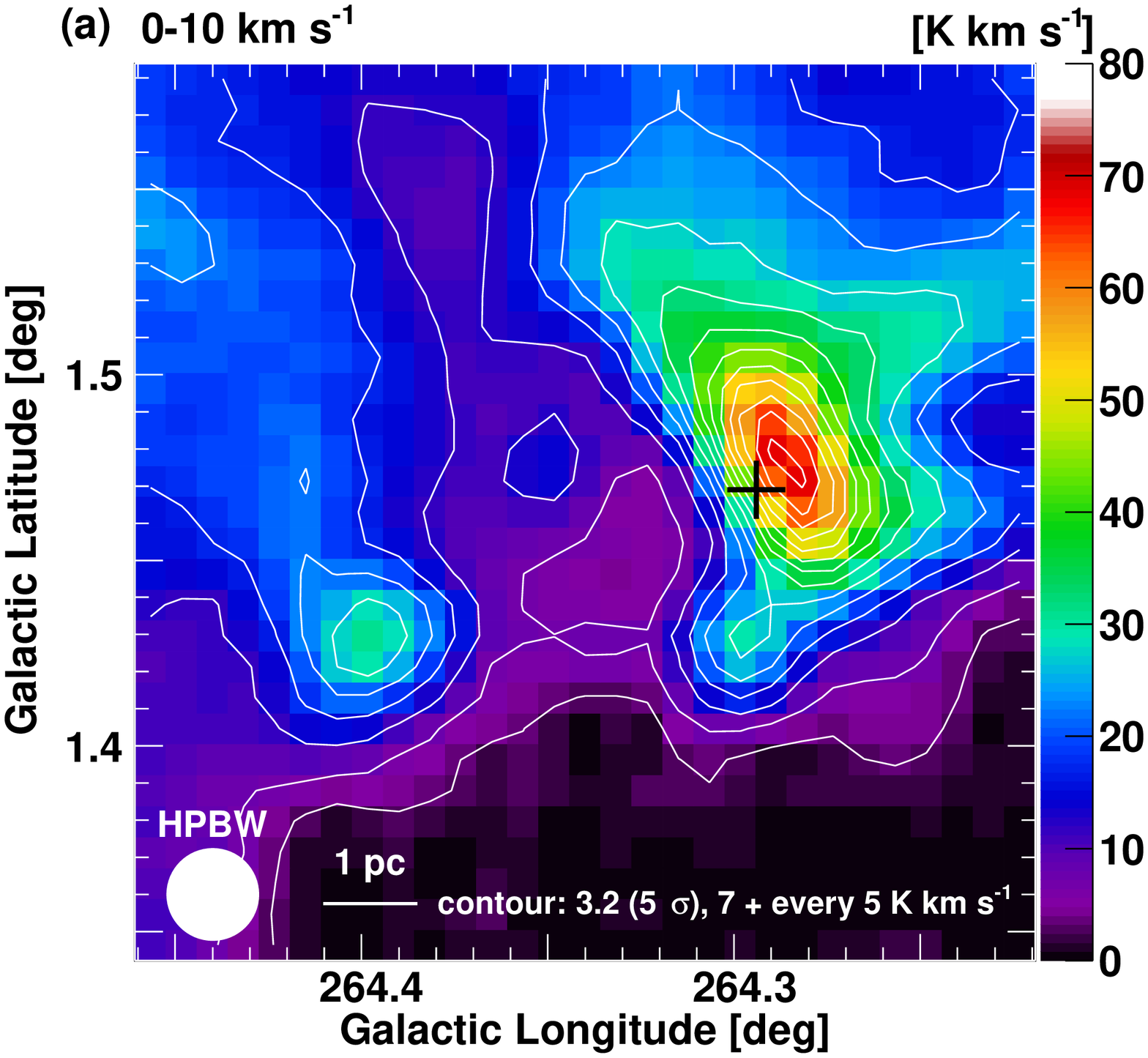}}}
   \end{center}
  \end{minipage} 
  \begin{minipage}{0.5\hsize}
   \begin{center}
    \rotatebox{0}{\resizebox{8cm}{!}{\includegraphics{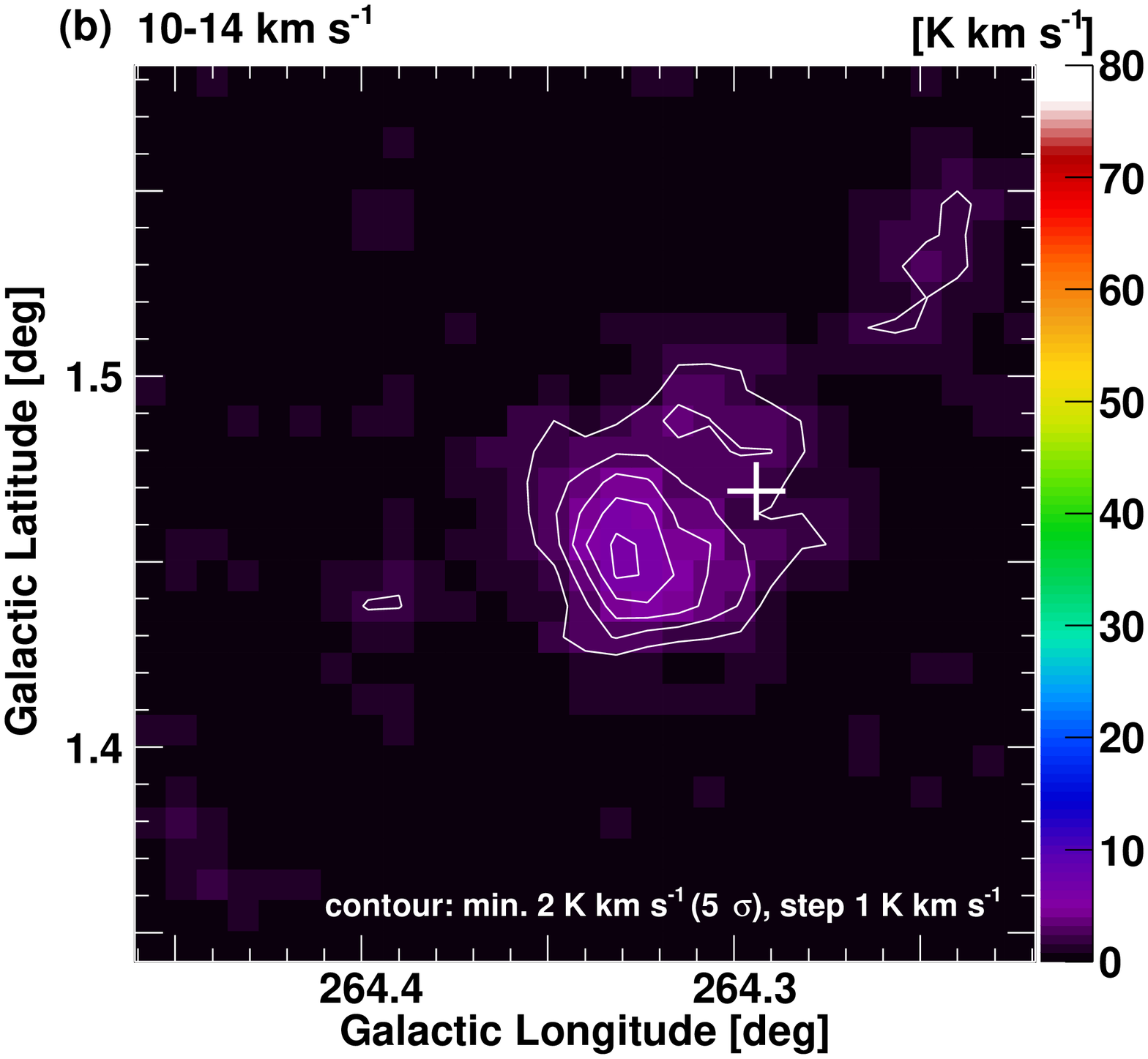}}}
   \end{center}
  \end{minipage} \\
   \begin{minipage}{0.5\hsize}
   \begin{center}
    \rotatebox{0}{\resizebox{8cm}{!}{\includegraphics{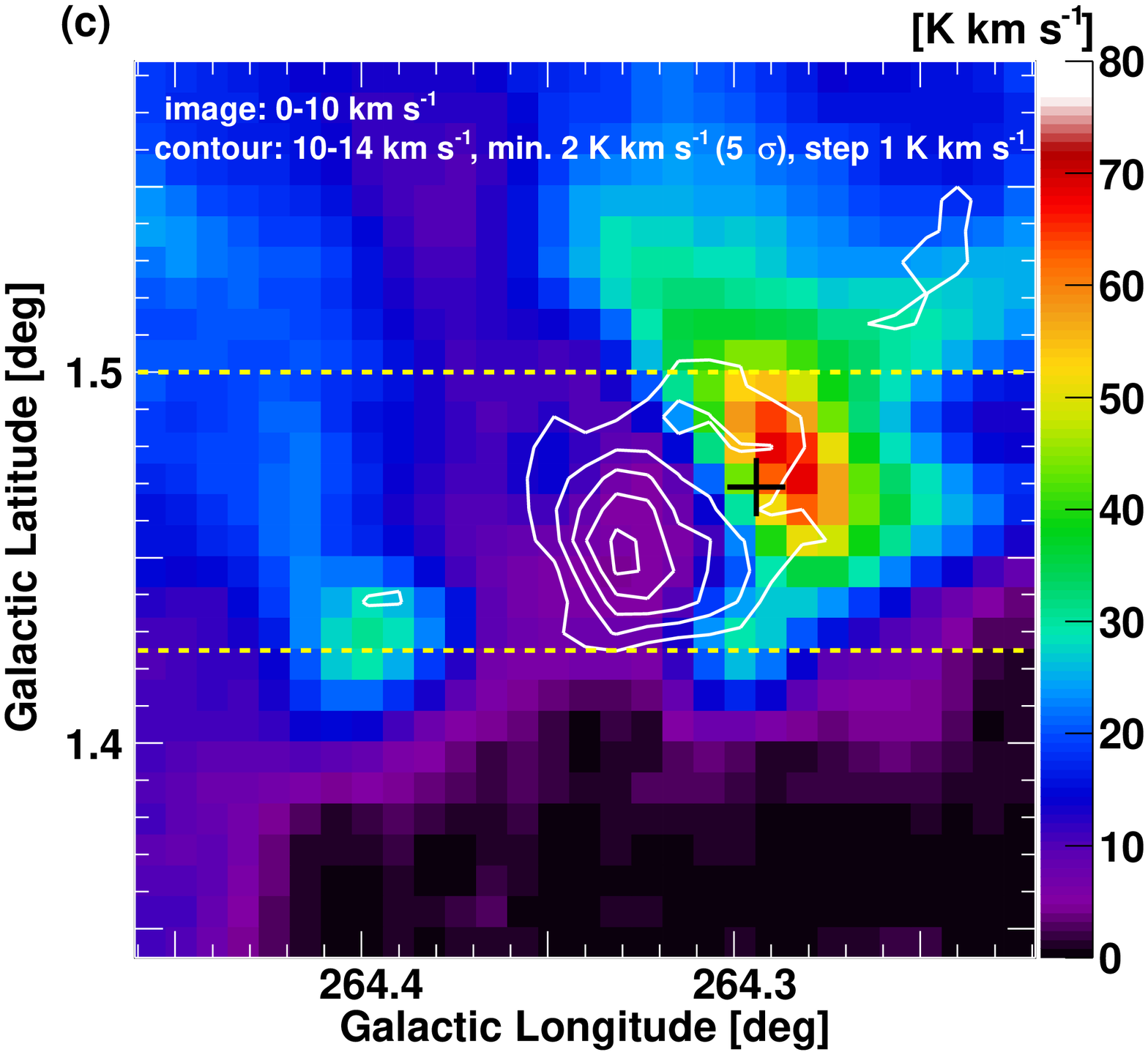}}}
   \end{center}
  \end{minipage} 
  \begin{minipage}{0.5\hsize}
   \begin{center}
    \rotatebox{0}{\resizebox{8cm}{!}{\includegraphics{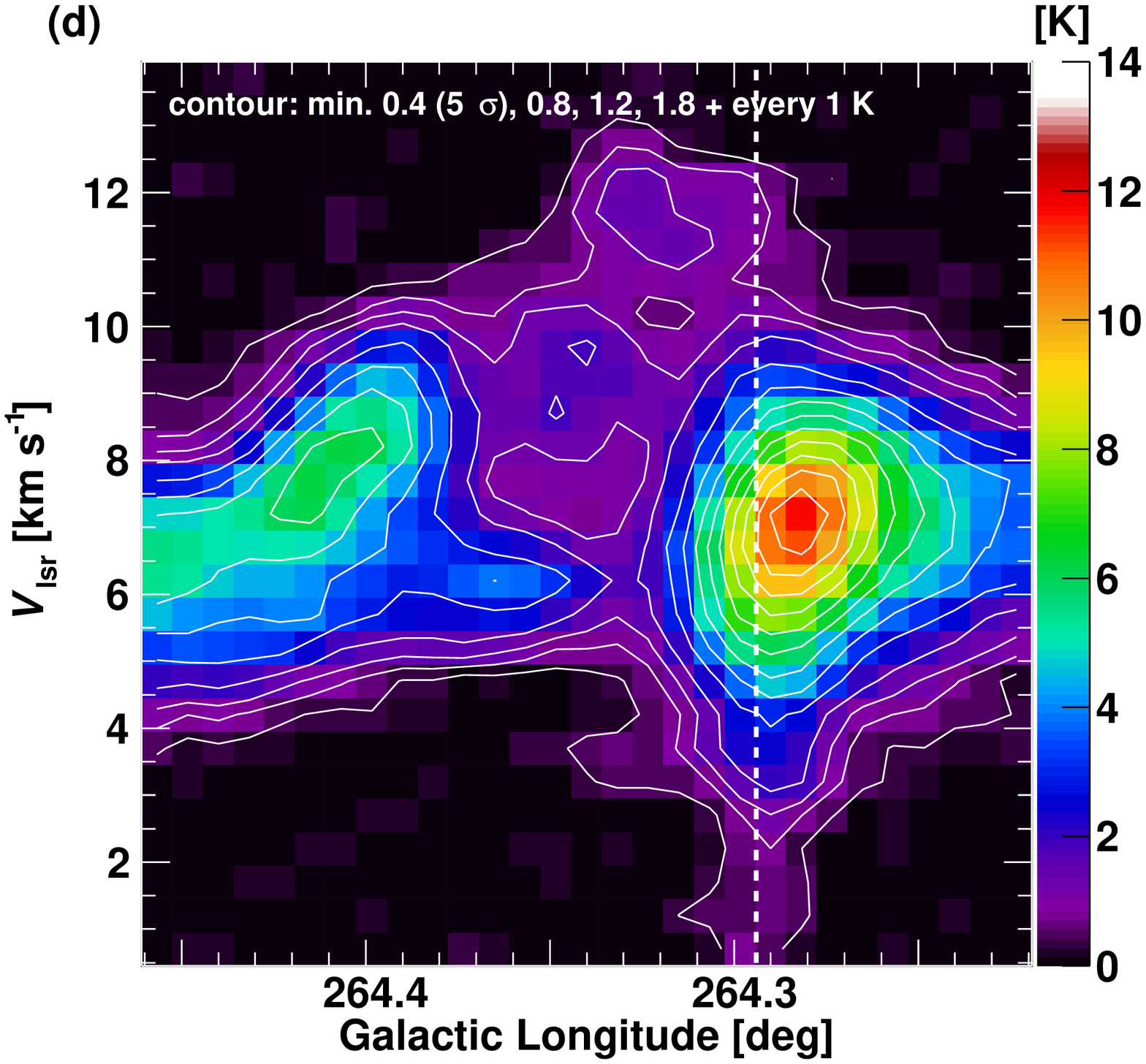}}}
   \end{center}
  \end{minipage} 
  \end{tabular}  
  \caption{Integrated intensity distributions of the $^{12}$CO $J=$2-1 data with the velocity range from (a) 0 to 10~km~s$^{-1}$, (b) 10 to 14~km~s$^{-1}$ and (c) a comparison of the two clouds. The cross indicates the position of O8.5V star. The yellow dashed lines indicate the integrated latitude range for the position-velocity diagram shown in the panel (d) and Figure~\ref{fig:12CO21_LV}~(a). (d): Longitude-velocity diagram with the integrated latitude range form $b =$ 1.425$^{\circ}$ to 1.5$^{\circ}$. The white dashed line indicates the position of O star.}
\label{fig:12CO21_LB}   
\end{figure}

In Figures~\ref{fig:12CO21_LB}(a) and~(b), we show the integrated intensity distributions of the blue and red clouds, respectively.
The panel (c) in the same figure compares positions of the two clouds represented by the contours (red cloud) overlaid on the image (blue cloud).
According to the peak velocities of each cloud (at ${\it V}_{\rm LSR}$~$\sim$7~km~s$^{-1}$ and $\sim$12~km~s$^{-1}$ for the blue and red clouds, respectively), these clouds have a relative velocity of $\sim$5~km~s$^{-1}$.
The exciting star is located between the two clouds, where the blue cloud shows a large gradient of the intensity, while the red cloud shows low significant emission with the intensity of $\sim$2~K~km~s$^{-1}$ ($\sim$5~$\sigma$). 
The northern part of the red cloud distributes likely to be enclosed by the blue cloud. 
The complementary distribution of the two clouds which harbors high-mass stars near the boundary between them is a characteristic structure recently has been found in several high-mass star-forming regions (e.g., \cite{Torii+15}; \cite{Fukui+17}; \cite{Torii+17}).
Figure~\ref{fig:12CO21_LB}(d) indicates a longitude-velocity diagram with the integrated latitude from $b =$ 1.425$^{\circ}$ to 1.5$^{\circ}$ as shown by the yellow dashed lines in the panel~(c).
In addition to the two intensity peaks included in the blue cloud (${\it V}_{\rm LSR} <$~10~km~s$^{-1}$), we confirmed the low significant ($\sim$5--15 $\sigma$) velocity structure corresponding to the red cloud in ${\it V}_{\rm LSR} >$~10~km~s$^{-1}$. 
The low intensity gas extended between the two peaks in the blue cloud corresponds to the bubble structure inside the infrared ring.
We found a wide velocity feature near the position of O star as shown by the white dashed line, which may suggest  outflows due to the YSOs around the massive stars. 

We estimated the molecular gas masses and the column densities of the blue and red clouds assuming that the intensity ratio between the $^{12}$CO $J=$2-1 and $J=$1-0 lines is 0.6, a typical intensity ratio obtained by the galactic plane survey performed by \citet{Yoda+10}, which gives an almost consistent result obtained by a recent theoretical study based on a hydrodynamical simulation \citep{Penaloza+17}.
With the assumptions of the distance 2.5~kpc toward RCW~34 \citep{Bik+10} and the CO-to-H$_{2}$ conversion factor,  $X_{\rm CO}=$ 1.0 $\times$10$^{20}$ H$_2$-molecule cm$^{-2}$ K$^{-1}$ km$^{-1}$~s (e.g., \cite{Okamoto+17}), the molecular masses for the blue and red clouds are derived to be $\sim$1.4 $\times$ $10^4$ $M_{\solar}$ and $\sim$600 $M_{\solar}$, respectively.
The peak values of the gas column density ($N_{\rm H_{2}}$) for the blue and red clouds are 1.1~$\times$~10$^{22}$ cm$^{-2}$ and 1~$\times$~10$^{21}$ cm$^{-2}$, respectively.
We also calculated the column density of the blue cloud to take into account the self-absorption effect often observed in the optically thick $^{12}$CO lines in high-density regions. 
Based on the model formula assuming the local thermodynamic equilibrium \citep{Nishimura+15}, we estimated optical depth of the $^{13}$CO $J=$2-1 line with the excitation temperature ($T_{\rm ex}$) obtained from the $^{12}$CO $J=$2-1 peak and derived the total molecular column density.
The typical optical depth of the $^{13}$CO $J=$2-1 line is $\sim$0.1 and the peak column density for the blue cloud is estimated to be $\sim$2.1~$\times$~10$^{22}$ cm$^{-2}$, which is consistent within the uncertainty by a factor of 2 compared to the value obtained with the $^{12}$CO $J=$2-1/$J=$1-0 intensity ratio.
The low CO intensity of the red cloud expects less self-absorption effect, and hence the column density of the red cloud would not change significantly.
Detailed studies of the column density and mass estimate are beyond scope of this paper.
Hereafter we use the column density and molecular gas mass estimated from the $^{12}$CO $J=$2-1/$J=$1-0 intensity ratio.

\clearpage
  
\subsection{$^{12}$CO $J=$3-2 Results}
\label{sec:12CO32} 

\begin{figure}[h]
 \begin{center}
 \centering
  \includegraphics[width=16.0cm]{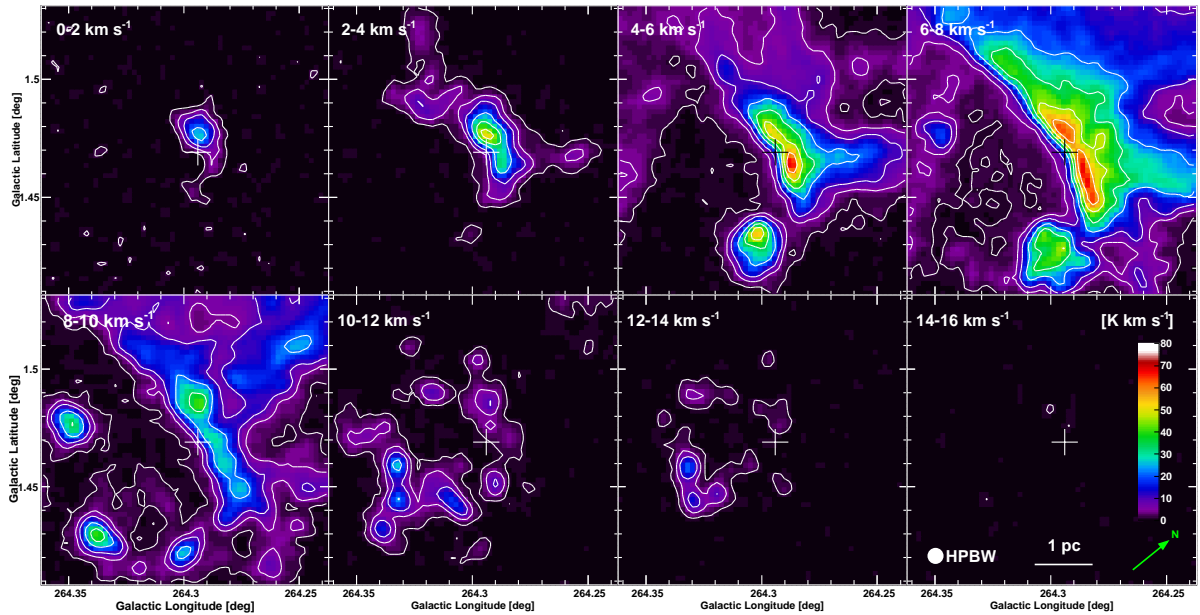}
  \end{center}
 \caption{Velocity channel distributions of the $^{12}$CO $J=$3-2 data. The contour levels are set to be 1.5, 5, and 10~K~km~s$^{-1}$, and to be followed by increasing 12~K~km~s$^{-1}$ step. The cross indicates the position of  O8.5V star.}
\label{fig:12CO32ChannelMap}  
\end{figure}

The $^{12}$CO $J=$3-2 data obtained with ASTE are useful to trace more excited gas properties in high-density clouds.
Figure~\ref{fig:12CO32ChannelMap} indicates the velocity channel distributions of the $J=$3-2 data separated into every 2 km~s$^{-1}$ with ${\it V}_{\rm LSR}$ from 0~to~14~km~s$^{-1}$.
Although the gas distribution is almost similar to that of $J=$2-1 data (see the velocity channel map in Figure~\ref{fig:12CO21ChannelMap}), the $J=$3-2 data with the higher angular resolution trace more tiny gas structures.
As the molecular gas distribution obtained by H$_2$ line observations (see Figure 4 in \cite{Bik+10}), the gas distributions in ${\it V}_{\rm LSR}$ from 2~km~s$^{-1}$ to 8~km~s$^{-1}$ for the blue cloud have mildly curved shape at the northern part to the O star, where the intensity peak at ${\it V}_{\rm LSR}$ $\sim$7~km~s$^{-1}$ is separated into two elongated structures. 
As traced by the $J=$2-1 velocity channel map in Figure~\ref{fig:12CO21ChannelMap}, low significant diffuse emission ($\sim$5 $\sigma$) are observed at ${\it V}_{\rm LSR}$~$\sim$7~km~s$^{-1}$ as well as the clumpy gas structures at ($l$, $b$) $\sim$ (\timeform{264D.35}, \timeform{1D.475}) and (\timeform{264D.34}, \timeform{1D.42}) with the sizes of $\sim$1~pc at ${\it V}_{\rm LSR}$~$\sim$9~km~s$^{-1}$ distributed in the bubble. 
While the $J=$2-1 data indicate that the red cloud (${\it V}_{\rm LSR}$ $>$ 10~km~s$^{-1}$) keeps a cloud morphology with almost symmetric structure in a significant detection level, the $J=$3-2 data in the low-density areas of the south-east part of the O star do not show significant $J=$3-2 emission (e.g., ($l$, $b$) $\sim$ (\timeform{264D.31}, \timeform{1D.47}) at ${\it V}_{\rm LSR}$~$\sim$11~km~s$^{-1}$).   
The non-detection of $J=$3-2 line might be due to low temperature or low density in the area compared with the critical density required for the $J=$3-2 transition.
The large beam size of the NANTEN2 $J=$2-1 data or insufficient sensitivity of the ASTE $J=$3-2 data possibly cause a difference of the CO distribution for the red cloud.
 
\clearpage

\subsection{$^{12}$CO $J=$3-2/$J=$2-1 Intensity Ratio}
\label{sec:LineRatio}

\begin{figure}[h]
 \begin{center}
 \centering
  \includegraphics[width=16.0cm]{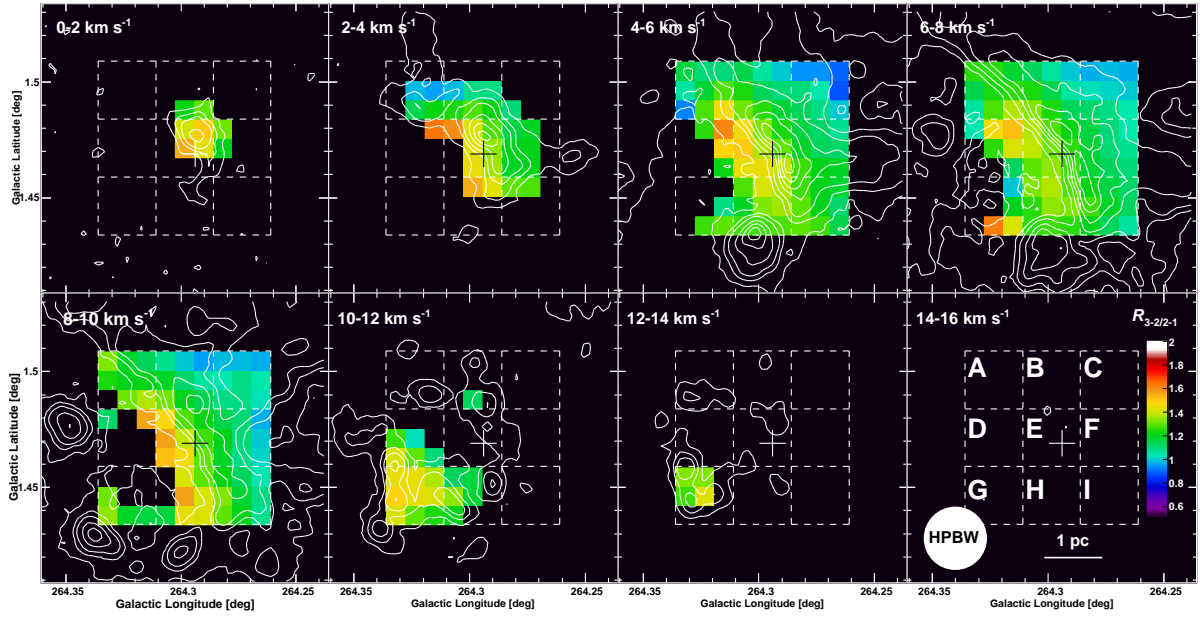}
  \end{center}
 \caption{Velocity channel distributions of the intensity ratio of $^{12}$CO $J=$3-2 to $J=$2-1. The contours indicate the integrated intensity of the $^{12}$CO $J=$3-2 line (as the same one in Figure~\ref{fig:12CO32ChannelMap}). The cross indicates the position of O8.5V star. Spectra for the areas of A--I are shown in Figure~\ref{fig:Spec_LocalArea}.}
\label{fig:12CO32to21ChannelMap}  
\end{figure}

Figure~\ref{fig:12CO32to21ChannelMap} indicates the velocity channel distributions representing the intensity ratio of the $^{12}$CO $J=$3-2 to $J=$2-1 (hereafter denoted by $R_{\rm 3-2/2-1}$). 
Taking intensity ratios between the different rotational transitions are useful to infer the gas properties of the exciting temperature or density. 
We need to evaluate carefully these ratios since the optically thick lines such as $^{12}$CO affect to lower the intensity especially in opaque regions.
To compare the intensities between the two lines, the $J=$3-2 data were smoothed to the NANTEN2 beam size of \timeform{90''}.
In the ratio map, we mask the three pixels from the edges, whose sizes are equivalent to the the beam diameter, because the correction of the beam size generates ambiguous uncertainties for the outermost pixels. 
In each velocity range, we also overlay contours to represent the integrated intensity of the $J=$3-2 line.
The obtained $R_{\rm 3-2/2-1}$ distribution exhibits that most of the molecular gas have high ratio ($\gtrsim$~1) throughout the all velocity ranges.
Using the three CO transition lines, we performed a Large Velocity Gradient analysis \citep{GoldreichKwan74} with assumptions of the abundance ratios, [$^{12}$CO]/[H$_2$] = 10$^{-4}$ (e.g., \cite{Frerking+82}; \cite{Leung+84}) and [$^{12}$C]/[$^{13}$C] $=$ 77 \citep{WilsonRood94}, and a typical velocity gradient roughly estimated from our $^{12}$CO $J=$2-1 observation, $dv/dr =$  (4 km s$^{-1}$)/(2 pc).
The results give a lower limit of $T_{\rm ex}\sim$200 K.
We confirmed $R_{\rm 3-2/2-1}$ $\gtrsim$ 1 in $T_{\rm ex}$ $>$200~K.
The optical depth of the $^{12}$CO lines is 2--8 within the typical H$_2$ density of the molecular clouds, 10$^2$--10$^4$ cm$^{-3}$, and the $J=$2-1 and 3-2 transitions have $T_{\rm ex}$ lower than the kinetic temperature ($T_{\rm k}$) (i.e., sub-thermal excitation), indicating that the observed brightness temperature ($T_{\rm b}$) for these lines are weaker than $T_{\rm k}$.
Even if we take into account the possible large optical depth in $^{12}$CO, the $R_{\rm 3-2/2-1}$ exhibits high values, supporting our result of the high intensity ratio shown in Figure~\ref{fig:12CO32to21ChannelMap}.

\begin{figure}[h]
 \begin{center}
 \centering
  \includegraphics[width=16.0cm]{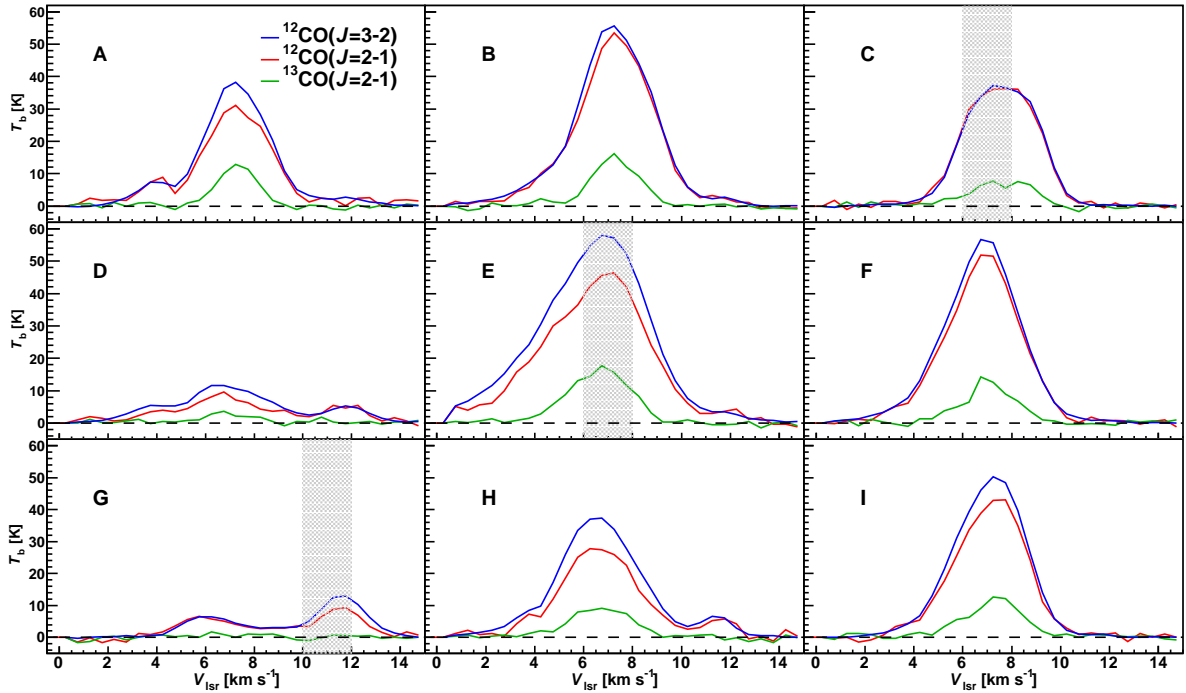}
  \end{center}
 \caption{$^{12}$CO $J=$3-2 (blue), $^{12}$CO $J=$2-1 (red) and $^{13}$CO $J=$2-1 (green) spectra with the brightness temperature ($T_{\rm b}$) for the areas of A--I (see Figure~\ref{fig:12CO32to21ChannelMap}). The shaded areas indicate the velocity range with the spectral peaks, which are discussed in the text.}
\label{fig:Spec_LocalArea}  
\end{figure}

To compare the line spectra among the different areas, we show the $^{12}$CO $J=$2-1, $J=$3-2 and $^{13}$CO $J=$2-1 spectra in Figure~\ref{fig:Spec_LocalArea} for the areas of A--I (shown in the bottom-right panel in Figure~\ref{fig:12CO32to21ChannelMap}).
These spectra represent sum of $T_{\rm b}$ within the 3 $\times$ 3 pixels.
The intensity peaks at ${\it V}_{\rm LSR} \sim$ 6--8 km~s$^{-1}$ in the blue cloud and ${\it V}_{\rm LSR} \sim$ 10--12 km~s$^{-1}$ in the red cloud indicate high $R_{\rm 3-2/2-1}$ as the following examples shown by the shaded areas: 
(i) The spectrum in area E containing the O-star position exhibits similar spectral shape but different intensities among the three lines, which gives a high $R_{\rm 3-2/2-1}$ ($\sim$1.5).
(ii) The area G located at $\sim$1--3 pc away from the O star, in which weak line emission from the red cloud is detected, also shows a high $R_{\rm 3-2/2-1}$ comparable to the area~E.
(iii) The high-density region in the blue cloud, area C, located oppositely to the area G around the O star, yields high $R_{\rm 3-2/2-1}$ value up to $\sim$1.0.
As shown by the integrated intensity map in Figure~\ref{fig:12CO21and13CO21}(b), significant $^{13}$CO $J=$2-1 emission from the area G is not detected.

\clearpage

\section{Discussion}
\label{sec:discussion}

\subsection{Properties of molecular clouds associated with the high-mass star}
\label{sec:GasProperties}

In the previous section, we found two molecular clouds separated in velocity with ranges of 0--10~km~s$^{-1}$ (blue cloud) and 10--14~km~s$^{-1}$ (red cloud), whose masses are $\sim$1.4 $\times$ 10$^{4}$~${\it M}_{\odot}$ and $\sim$600~${\it M}_{\odot}$, respectively.
We summarize several pieces of evidence of physical associations between the two clouds and the massive stars as the followings.

\begin{enumerate}

\item The blue and red clouds show a complementary distribution. While the blue cloud with an arc-shaped structure distributes  along the infrared 8 $\micron$ ring, the red cloud distributes just likely to cover the northernmost of the bubble, which consists of the diffuse gas emission and clumpy gas structures with the sizes of $\sim$1 pc. The O / early B stars are located near the boundary between the two clouds, where the blue cloud is possibly highly compressed due to the feedback from the massive  stars, as suggested by its mildly curved-shape in the neighborhood of the O star. The diffuse gas component near the massive stars in the red cloud is probably disrupted by the strong UV radiation.

\item From the position-velocity diagram, we confirmed each velocity of the two clouds, as well as the bubble structure  sandwiched by the two peaks of the blue cloud. We will discuss this velocity feature in a comparison with a simulation result in Section~\ref{sec:CompSimulation}.  

\item The line intensity ratio of the $^{12}$CO $J=$3-2 / $J=$2-1 in the blue and red clouds shows high values ($\gtrsim$~1.0), possibly due to heating by the UV radiation from the massive stars, suggesting physical associations between the two clouds and the high-mass stars.

\end{enumerate}

\begin{figure}[h]
 \begin{center}
 \centering
  \includegraphics[width=16.0cm]{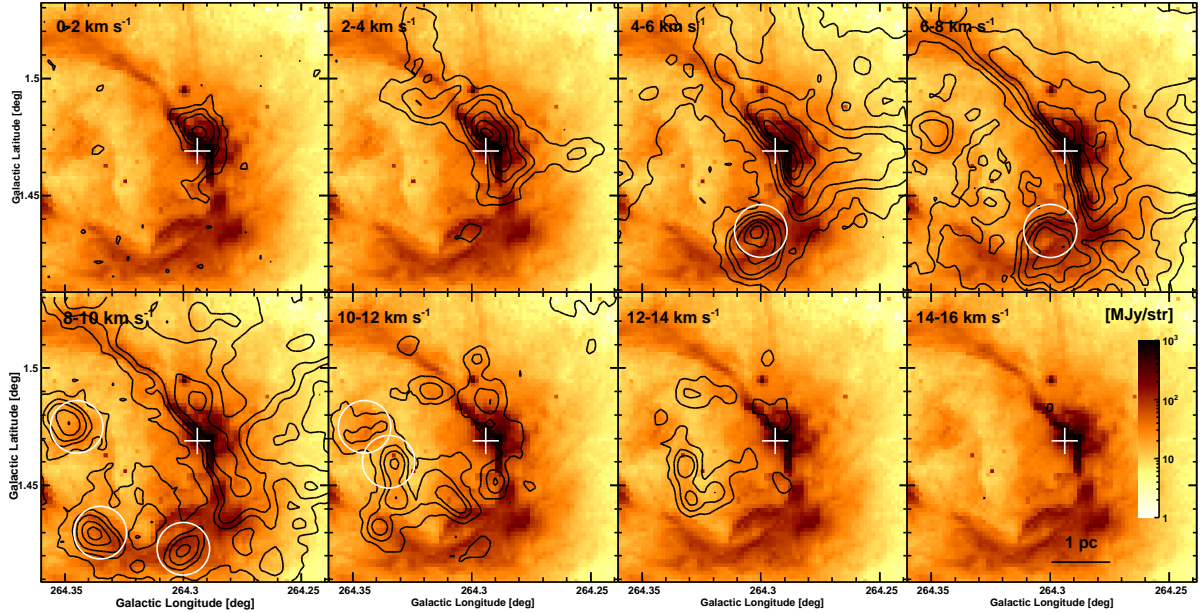}
  \end{center}
 \caption{A comparison of the {\it Spitzer} 8 $\micron$ (image) and $^{12}$CO $J=$3-2 (contour) distributions in each velocity range of 2~km~s$^{-1}$. The CO contour levels are set to be 1.5, 5, and 10~K~km~s$^{-1}$, and followed by increasing 12~K~km~s$^{-1}$ step. The cross indicates the position of O8.5V star. The white circles indicate the clumpy gas regions where the correlation between the CO and 8~$\micron$ emission are clearly seen.} 
\label{fig:12CO32ChannelMap_8um}  
\end{figure}

In Figure~\ref{fig:12CO32ChannelMap_8um}, we compare the distribution of $^{12}$CO $J=$3-2 line with that of the infrared 8~$\micron$ in a velocity channel map.
The 8~$\micron$ emission mainly trace the PAH, which are probably created by the photodissociation induced by the UV radiation from the massive stars. 
As shown by the white circles in the figure, we found good correspondence between the clumpy gas structures and the 8~$\micron$ distributions both for the blue and red clouds.
In each clump, the 8~$\micron$ emission on the side facing to the exciting stars tend to be higher as compared to the opposite side.
These results indicate that the molecular clouds distributed in the bubble are strongly affected by the UV radiation in a far away distance at least $\sim$3 pc, supporting physical associations between the molecular clouds and the massive stars.

\subsection{Possible scenarios to form the gas and stellar distributions in RCW~34}
\label{sec:PreviousScenarios}

A triggered mechanism of the massive-star formation in the H{\sc ii} region is generally explained by the collect \& collapse \citep{ElmegreenLada77}.
An expanding motion of the ionized gas due to the feedback from the massive star sweeps up the surrounding material between the ionization front and the shock front (e.g., \cite{HosokawaInutsuka06}).
The collected material forming a dense molecular layer becomes gravitationally unstable and collapses to form  next-generation of stars in individual fragmentations (e.g., \cite{Whitworth+94}). 
There are several candidates for the high-mass star-formation site on the borders of the H{\sc ii} region, which can be explained by the collect \& collapse (e.g., \cite{Deharveng+08}; \cite{Zavagno+10}).
Assuming an ideal uniform medium to form a massive star via the collect \& collapse, star-formation processes should be observed regardless of the directions, which yields a symmetric gas distribution (e.g., Figure 1 in \cite{Deharveng+10} and Figure~8 in \cite{Torii+15}) to form next-generation of stars.
In practice, such a concentric gas distribution may not be observed because of gas disruption due to the radiative heating and the projection effect along the line of sight.
Initial condition to trigger forming massive stars can also change the observed gas distributions. 
However, as suggested by \citet{Bik+10}, more clumps should be observed in the southern area of the O star, if the collect \& collapse can be applied to RCW~34.
YSOs such as the class 0/I and I/II objects corrected to the gas distribution would be found in the inner bubble (see Figure~\ref{fig:ThreeCompositeColorMap}).

\citet{Bik+10} proposed another scenario, a sequential star formation from the south to north, which eventually gives rise to the formation of massive stars including the YSOs such as the class 0/I objects located on the northern part of the bubble.
However, it is unclear that there is an age gradient across the whole bubble because the formation time scale of the class II objects distributed in the bubble has large uncertainty (a few Myr; \citet{Bik+10} and references therein).
In addition, the class 0/I and I/II objects located around ($l$, $b$) $\sim$ (\timeform{264D.33}, \timeform{1D.43}) are hard to explain with the scenario of sequential star-formation from the south to north. 
The class II objects are likely to be spread out in the bubble extensively rather than clustering in one zone.
We infer that the formation process of the class II objects in the bubble are not strongly related to the past formation history of the massive stars and the ring-like gas distribution.

As an alternative interpretation, we propose a scenario for the formation of massive star and the ring-like structure in RCW~34 via the cloud-cloud collision, which can explain properties of the molecular clouds summarized in Section~\ref{sec:GasProperties}.
An accidental supersonic collision between the blue and red clouds induces the formation of cloud cores near the boundary of the clouds as a consequence of the shock compression, giving a trigger of a high accretion rate to form the high-mass stars.  
In the center of the blue cloud, a large cavity is created by the collisional effect, resulting in the complementary distribution of the two clouds.
The molecular gas around the O star are heated by the strong UV radiation and ultimately disrupted, where the CO emission becomes weak or cannot be observed.
Molecular clouds accompanying high-mass stars have been found recently in several super star clusters and H{\sc ii} regions, which can be explained by the cloud-cloud collision (\cite{Furukawa+09}; \cite{Ohama+10}; \cite{Fukui+14}; \cite{Torii+15}; \cite{Fukui+16}; \yearcite{Fukui+17}; \cite{Torii+17}). 
The cloud-cloud collision model can be also applied to the other high-mass star-forming regions in the Vela Molecular Ridge (RCW~38, \cite{Fukui+16}; RCW~32, Enokiya et al. 2017 in prep; RCW~36, \cite{Sano+17}).
Among these candidates summarized in \citet{Fukui+17}, RCW~34 with a single O star is characterized by the two clouds with relatively lower mass.

In establishing the model of high-mass star-formation, constraint on the initial condition to trigger the star-forming process is one of a significant subject. Although the major theoretical models ``core accretion'' and ``competitive accretion'' considering the turbulent effect are commonly thought to be promising pictures to explain the high-accretion rate to form massive stars (e.g., \cite{McKeeTan03}), the gas column density in RCW~34 obtained in this study ($\sim$4$\times$10$^{-2}$ g cm$^{-2}$ for blue cloud) does not satisfy the threshold of the column density to halt the fragmentations (1 g cm$^{-2}$; \cite{KrumholzMcKee08}). It is difficult to convince these models observationally in terms of the required initial condition. The model of collect \& collapse including the sequential star-formation suggested by \citet{Bik+10} described above does not give observational constraints on the triggering process on the onset of the sequential star-formation. On the other hand, the model of cloud-cloud collision based on the gravitationally unbound system can satisfactorily explain in RCW 34 the triggering process to form the massive stars by likely amplifying the magnetic field and the turbulent velocity for high-mass accretion rate \citep{InoueFukui13}.

The velocity channel map in Figure~\ref{fig:12CO21ChannelMap} exhibits lobe-like gas distributions with the highly blue-shifted ($V_{\rm lsr} \sim$~0--4 km~s$^{-1}$) and the red-shifted ($V_{\rm lsr} \sim$~10--14 km~s$^{-1}$) components toward the northern and southern directions, respectively, indicating a possibility that the gas distribution in RCW~34 is formed by molecular outflow from the star-forming region around the O star instead of the collision of the two clouds.
This is supported by the wide velocity feature found in the position-velocity diagram in Figure~\ref{fig:12CO21_LB}(d). 
As shown in Figure~\ref{fig:Spec_LocalArea}, most of the $^{12}$CO spectra indicate possible blue-shifted lobe (for areas A, B, D, E, F, H and I), whereas the spectrum for area G does not have significant emission of the blue-shifted lobe. 
The spectrum in area G shows the significant red-shifted lobe, whereas  another emission peak at $V_{\rm lsr} \sim$6 km~s$^{-1}$ is detected. 
Such spectral feature with the velocity jump is difficult to understand with the molecular outflow from the massive star-forming region around the O star;
the significant emission connecting the red-shifted lobe (included in the red cloud) and the $\sim$6~km~s$^{-1}$ peak (included in the blue cloud) can be rather attributed to a bridging feature often observed in candidates of the cloud-cloud collision.
Hereafter we discuss a possibility that the formations of the O star and the ring-like gas distribution in RCW~34 are via a collision between the two clouds.

\subsection{Velocity Structure Compared with a Simulation Result}
\label{sec:CompSimulation} 
 
\begin{figure}[h]
 \begin{tabular}{cc}
  \begin{minipage}{0.5\hsize}
   \begin{center}
    \rotatebox{0}{\resizebox{8cm}{!}{\includegraphics{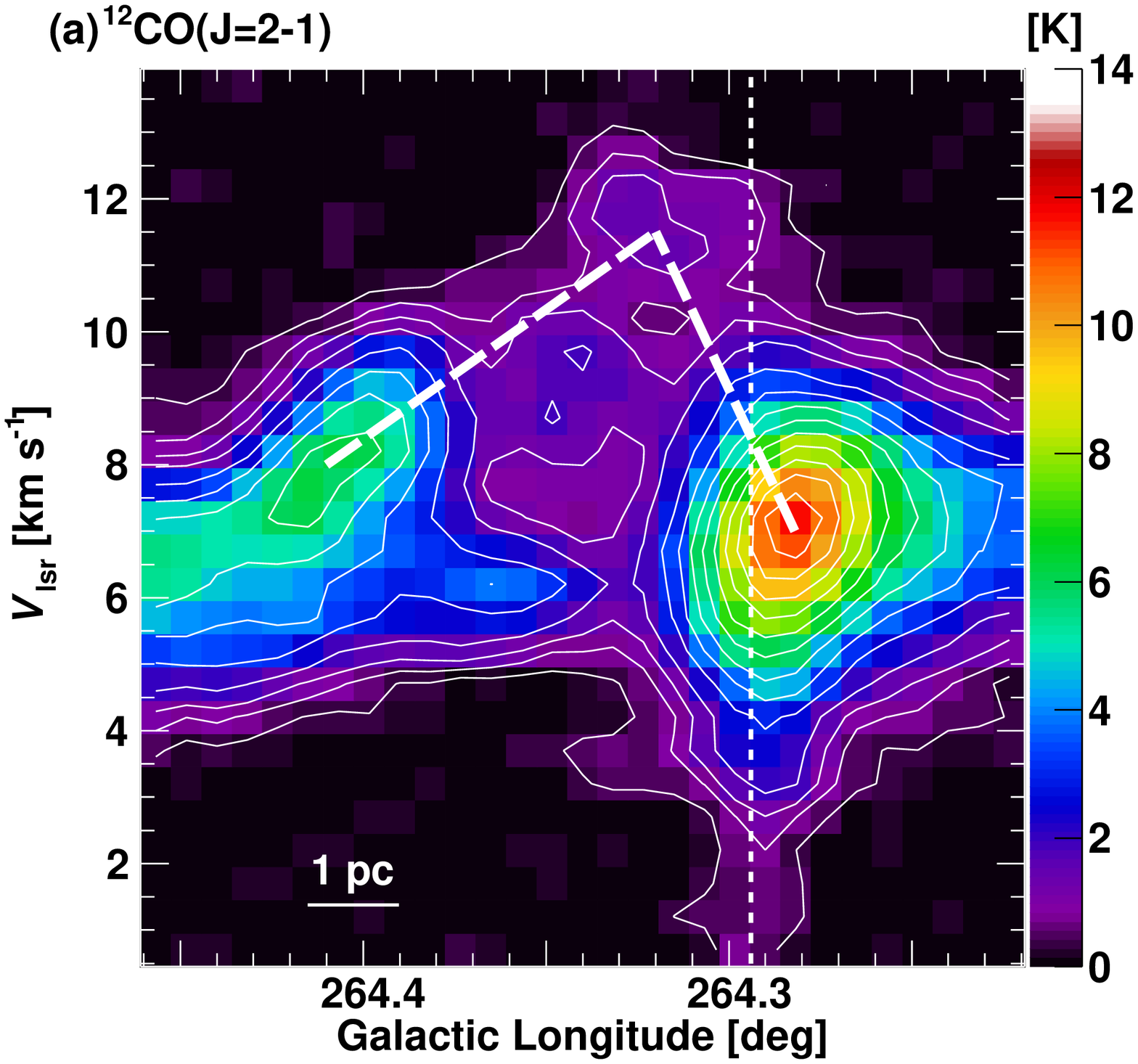}}}
   \end{center}
  \end{minipage} 
  \begin{minipage}{0.5\hsize}
   \begin{center}
    \rotatebox{0}{\resizebox{8cm}{!}{\includegraphics{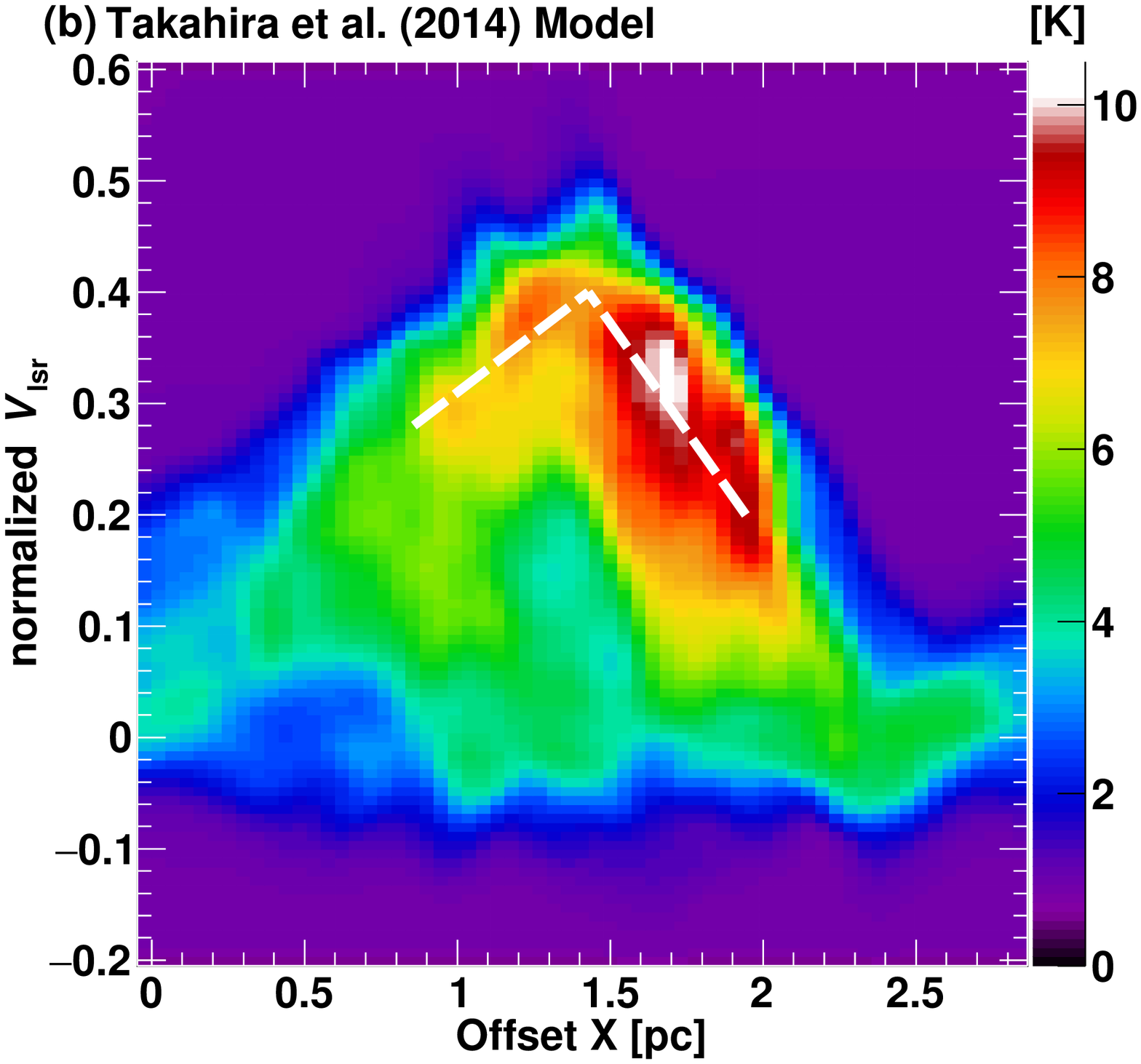}}}
   \end{center}
  \end{minipage} 
  \end{tabular}  
  \caption{(a): Longitude-velocity diagram of the $^{12}$CO $J=$2-1 data integrated from $b =$ 1.425$^{\circ}$ to 1.5$^{\circ}$ (same as Figure~\ref{fig:12CO21_LB}~(d)). The white thick dashed line represents the inverted v-shape structure which connects the peaks of the blue and red clouds. The white thin dashed line indicates the position of O star. (b): Position-velocity diagram obtained by a synthetic $^{12}$CO $J=$1-0 observation (\cite{Haworth+15a}; \yearcite{Haworth+15b}) using the collision model provided by \citet{Takahira+14}. The image is smoothed to be the size of spatial and velocity resolutions in the $^{12}$CO $J=$2-1 observation. The integrated Y axis is from $-$1 pc to 1 pc with reference to the collision axis. The velocity axis is normalized to the initial colliding velocity 10~km~s$^{-1}$. The white dashed line indicates an auxiliary line to represent the inverted v-shape structure.}
\label{fig:12CO21_LV}   
\end{figure}

Figure~\ref{fig:12CO21_LV}(a) indicates the longitude-velocity diagram of the $^{12}$CO $J=$2-1 data (same as Figure~\ref{fig:12CO21_LB}(d)), in which the two lines connecting the intensity peaks in the blue and red clouds are overlaid.
A velocity gradient characterized by an inverted v-shape structure emerges, implying a possibility that the central cavity was formed by crashing the red cloud into the blue cloud with an angle in the direction to the lower Galactic longitude to the line of sight.

We compared this velocity feature with a numerical simulation of the cloud-cloud collision, in which the clouds with a different size of Bonnor-Ebert spheres \citep{Bonnor56} are assumed \citep{Takahira+14}.
In this comparison, we do not compare the $^{12}$CO $J=$3-2 data, because the $J=$3-2 observation does not cover the whole clouds.
In the model calculation, turbulent effect is given to the clouds, although any radiative feedback from the massive stars is not considered.
Assuming the initial collision velocity 10~km~s$^{-1}$ with a viewing angle 45$^{\circ}$ to the line of sight, we analyzed data obtained by a synthetic $^{12}$CO $J=$1-0 observation (\cite{Haworth+15a}; \yearcite{Haworth+15b}) based on the collisional model provided by \citet{Takahira+14}.
We note that physical conditions of the colliding clouds in this simulation do not match our observational study, but a comparison with the simulation result is useful to investigate qualitatively how the velocity distribution is observed by the collisional effect. 
Figure~\ref{fig:12CO21_LV}(b) indicates a snapshot of the position-velocity diagram at 1.6 Myr after the collision started, when the relative speed was decreased to 7~km~s$^{-1}$ \footnote{The collisional interaction decelerates the relative speed depending on time (see Figure 3 in \cite{Haworth+15b})}.
We confirmed gas distribution having the inverted v-shaped and cavity-like structures similar to our observational result, which suggests a past collisional event in RCW~34 between the blue and red clouds.
The simulation result traces a highly-compressed region shown by the elongated high intensity feature, which corresponds to the bridging site connecting the two clouds in velocity.
Turbulent gas motions are expected there, possibly leading to the high-mass star-formation.
Our observational results, however, are difficult to verify tiny gas structures due to the poor spatial and velocity resolutions.
Future follow up observations by ASTE toward the southern area will provide the velocity feature of the whole clouds and ALMA observations of the colliding area are expected to resolve filamentary gas structures associated with the turbulent gas motions suggested by a numerical simulation (\cite{InoueFukui13} and \cite{Inoue+17}), which will be useful to finally probe that cloud-cloud collision is a relevant process in the formation of high-mass stars.  
 
\subsection{Collisional Time Scale}
\label{sec:CollisionalTimeScale} 

\begin{figure}[h]
 \begin{center}
 \centering
  \includegraphics[width=8.0cm]{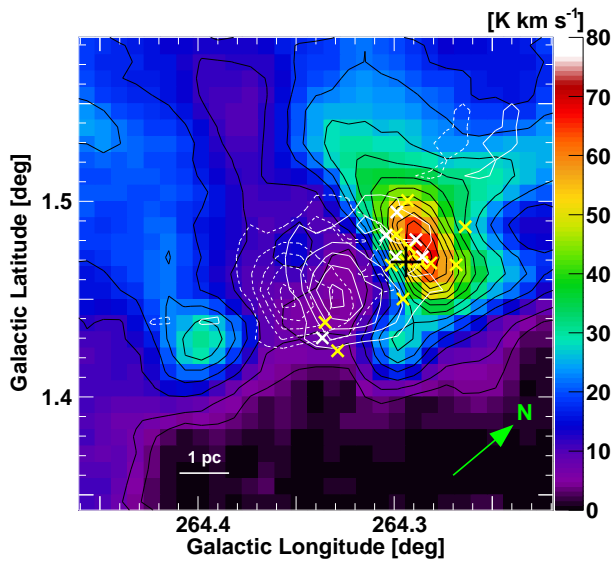}
  \end{center}
 \caption{Integrated intensity map of the $^{12}$CO $J=$2-1 data for the blue cloud (the same as Figures~\ref{fig:12CO21_LB}~(a) and (c)), in which two components of the red cloud are shown by the white contours: the solid line indicates the position of current observed cloud and the dashed line indicates the position with a 1 pc displacement toward the positive longitude direction. The black cross indicates the O star and yellow and white crosses indicate the class 0/I and I/II objects \citep{Bik+10}, respectively.}
\label{fig:12CO21_LV_Disp}  
\end{figure}

Figure~\ref{fig:12CO21_LV_Disp} indicates the integrated intensity distribution of the $^{12}$CO $J=$2-1 data for the blue cloud (same as Figures~\ref{fig:12CO21_LB}~(a) and (c)), overlaying the intensity of the red cloud by the white contours.
The dashed contours indicate the intensities for the red cloud applying a displacement of 1 pc toward the positive longitude direction. 
The displacement makes the red cloud cover the overall bubble, showing a further morphological correspondence between the blue and red clouds.
If the cloud-cloud collision model can be applied to these clouds, an existence of the cavity whose size is the red cloud is probable, and a movement of the red cloud toward the north direction is expected.
The morphological correspondence with the displacement implies that the collision started from there and the formation of O star was triggered by a continuous high-compression at the shock front.
We therefore adopted the size of displacement 1~pc to estimate the collisional time scale.  
If we apply tentatively 45$^{\circ}$ for the viewing angle of the collision to the line of sight, the distance between the two clouds is $D=\ $1/sin(45$^{\circ}$) (pc) and the velocity difference estimated from the observed relative velocity is $V=\ $5/cos(45$^{\circ}$) (km~s$^{-1}$).
This inclination angle roughly matches the collisional direction suggested from the longitude-velocity diagram discussed in Section~\ref{sec:CompSimulation}.
The collisional time scale can be estimated to be $D$/$V$$\sim$1.5 Myr.
If we take into account variable viewing angles of the collision to the line of sight from 30$^{\circ}$ to 60$^{\circ}$, the collisional time scale is derived to be 0.1--0.35 Myr.
We note that actual collisional time scale will be more shorter since the initial velocity is expected to be higher and reduced to the current velocity due to the deceleration effect.
After the collision started, continuous high-pressure had been driven between the clouds, which gave a trigger to form the high-mass star; the time scale forming the O star is inferred to be $\lesssim$~0.2~Myr.

The collisional time scale $\sim$0.2 Myr to form the H{\sc ii} region RCW~34 is much shorter than the age of 2$\pm$1 Myr, which is estimated by the population of pre-main-sequence (PMS) stars distributed in high-density regions around the massive stars  \citep{Bik+10}. 
Meanwhile, the author noted a possibility that these PMS stars are likely much younger because outflows are detected, possibly originating from the YSOs such as the class 0/I objects distributed in the neighborhood of the massive stars (see Figure~\ref{fig:12CO21_LV_Disp}). 
In generally, the time scale of the class I objects is thought to be $\sim$ a few 10$^{5}$ years (e.g., \cite{Whitney+03}), which is almost consistent with that of O-star formation inferred by the cloud collision.
The formation of class~0/I and I/II objects located in the southern area is difficult to explain by the sequential star formation from the south to north suggested by \citet{Bik+10}.
If the cloud-cloud collision model can be applied to the low-mass star formation \citep{Nakamura+12}, the existence of local class 0/I and I/II objects is understood with the same formation time scale of the other class~0/I and I/II objects.
In terms of the formation time scale of the class~II objects (a few Myr; \citet{Bik+10} and references therein) and their distributions extensively spread out in the whole bubble, these YSOs might have been formed prior to the cloud collision.

\subsection{Comparison with Other Star Forming Region}
\label{sec:CompOtherSFR} 

Finally, we compared properties of molecular clouds in RCW~34 with those of other candidates of the cloud-cloud collision, RCW~120 and M20 (\cite{Torii+15}; \yearcite{Torii+17}), which accompany a single O star in their H{\sc ii} regions, and the super star cluster, NGC~3603 and RCW~38, having a few tens of massive stars (\cite{Fukui+14}; \yearcite{Fukui+16}). 
Table~\ref{table:CompCCCcandidates} summarizes physical parameters of these candidates.
We note that the uncertainties due to utilizing different CO lines as the gas tracer, angular resolutions, and the assumptions of $X_{\rm CO}$ among these high-mass star-forming regions do not affect this discussion of the scale of column density in the order of magnitude.
\citet{Fukui+17} suggested that the molecular column density of the colliding clouds can be an important parameter to determine the number of O stars.
Whereas NGC~3603 and RCW~38 have high column density up to 10$^{23}$~cm$^{-2}$, other candidates to form a single O star show a lower column density with the range of 10$^{21}$ to $\lesssim$~10$^{22}$~cm$^{-2}$.
The present study of RCW~34 also shows a similar diagnostic that the clouds with the column density $\sim$10$^{22}$~cm$^{-2}$ can form a single O star.  
Compared to the other two candidates of RCW~120 and M20 accompanying a single O star, RCW~34 has a large difference of the column density between the two clouds. 
The formation of O star might be triggered if one of the parental clouds has enough large ($\gtrsim$~10$^{22}$~cm$^{-2}$) column density. 
We also note that the time scale forming high-mass stars are almost consistent between RCW~34 and RCW~38 in spite of the large different number of massive stars, which may indicate physically associated star-formation history via the cloud-cloud collision in the Vela Molecular Ridge.
Similar time scale of the cloud collision to trigger the formation of O stars derived in RCW~36 \citep{Sano+17} supports this indication.
We need molecular observations toward other star-forming regions to reveal properties of clouds related to the massive stars, which may establish the scenario to form the high-mass stars triggered by the cloud-cloud collision.

\begin{table}[h]
 \caption{\normalsize{Comparison of the molecular cloud properties of high-mass star-forming regions among the several candidates of the cloud-cloud collision}} 
 \label{table:CompCCCcandidates}
  \begin{center}
   \begin{tabular}{ccccccc} \hline\hline
   \makebox[3em][c]{Name} &
   \makebox[5em][c]{Number of} &
   \makebox[3em][c]{Cloud} &
   \makebox[6em][c]{Molecular} &
   \makebox[3em][c]{Relative} &
   \makebox[4em][c]{Collision} &
   \makebox[6em][c]{References} \\ 
   & O stars & Mass & Column Density & Velocity & Time Scale & \\ 
   & ($\times$ 10$^{3}$ ${\it M}_{\odot}$) & ($\times$ 10$^{22}$ cm$^{-2}$) & (km s$^{-1}$) & (Myr) & \\ \hline
   RCW 38 & $\sim$20 & (30, 2) & (10, 1) & 12 & $\sim$0.1 & \citet{Fukui+16} \\
   NGC 3603 & $\sim$30 & (70, 10) & (10, 1) & 15 & $\sim$1 & \citet{Fukui+14} \\ \hline
   RCW 120 & 1 (O8V or O9V) & (50, 4) & (3, 0.8) & 20 & $\sim$0.2 & \citet{Torii+15} \\ 
   M20 & 1 (O7.5) & (1, 1) & (1, 1) & 7.5 & $\sim$0.3 & \citet{Torii+17} \\
   RCW 34 & 1 (O8.5V) & (14, 0.6) & (1.1, 0.1) & 5 & $\sim$0.2 & This study \\ \hline  	
   \end{tabular}
  \end{center}
\end{table}

\section{Conclusion}
\label{sec:conclusion}

Using the $^{12}$CO and $^{13}$CO~$J=$2-1, and $^{12}$CO $J=$3-2 data obtained by the NANTEN2 and ASTE telescopes, we have studied molecular gas distribution toward the H{\sc ii} region RCW~34, in which a single O star and two early B type stars are located.
The main conclusions are summarized below.

\begin{enumerate}
  \item 
  We found two clouds separated in velocity with ranges of 0--10~km~s$^{-1}$ (blue cloud) and 10--14~km~s$^{-1}$ (red cloud), whose masses are $\sim$1.4 $\times$ 10$^{4}$~${\it M}_{\odot}$  and $\sim$600~${\it M}_{\odot}$, respectively.
  The blue cloud distributes likely to trace the ring-like structure observed in the infrared wavelengths. 
  The red cloud, which has not been recognized in previous observations, consists of the diffuse gas emission and clumpy gas structures, just to cover the northernmost of the bubble.   
  The blue and red clouds show a complementary distribution.
  The high-mass stars with the spectral types of O8.5V and early B are located near the boundary between the two clouds.
  \item 
   The blue cloud is likely to be compressed by the feedback from the O star. 
   The weak CO emission in the red cloud indicates disruptions by the ionization effect from the massive stars. 
   The line intensity ratio of $^{12}$CO $J=$3-2 / 2-1 shows high values ($\gtrsim$~1.0), possibly due to heating by the UV radiation. 
  The infrared 8 $\micron$ emission shows good spatial correspondence with the molecular clouds.
  These results suggest physical associations between the two clouds and the high-mass stars.
  \item
  The position-velocity diagram of $^{12}$CO~$J=$2-1 data shows an inverted v-shape structure, which is similar distribution obtained by a synthetic observation with a numerical simulation. 
  We need follow up observations by ASTE and ALMA to discuss the velocity feature more finely resolved scales in the whole clouds. 
  If we apply a 1~pc displacement for the red cloud, the morphological correspondence between the two clouds becomes more pronounced.
  Assuming the collision at a viewing angle 45$^{\circ}$ to the line of sight, the collisional time scale is estimated to be $\sim$0.2 Myr, which indicates that the high-mass stars and the surrounding structure in RCW~34 were formed in a much short time scale $\lesssim$~0.2 Myr. 
  Compared to other massive star-forming regions identified as candidates of the cloud-cloud collision, RCW~34 is characterized by a high contrast in the molecular column density (or mass) between the two clouds.
  The number of O stars formed by the triggering via the cloud-cloud collision may relate to their column densities.
  
\end{enumerate}

\begin{ack}
We would like to thank Arjan Bik for providing information of the YSOs distributed in RCW~34.
NANTEN2 is an international collaboration of ten universities, Nagoya University, Osaka Prefecture University, University of Cologne, University of Bonn, Seoul National University, University of Chile, University of New South Wales, Macquarie University, University of Sydney, and Zurich Technical University.
The ASTE telescope is operated by National Astronomical Observatory of Japan (NAOJ).
\end{ack}


\begin{thebibliography}{}

\bibitem[Anathpindika(2010)]{Anathpindika10} Anathpindika, S. V. 2010, MNRAS, 405, 1431
\bibitem[Bik et al.(2010)]{Bik+10} Bik, A., Puga, E., Waters, L. B. F. M., et al. 2010, ApJ, 713, 883
\bibitem[Bonnor(1956)]{Bonnor56} Bonner, W. B. 1956, MNRAS, 116, 351
\bibitem[Deharveng et al.(2008)]{Deharveng+08} Deharveng, L., Lefloch, B., Kurtz, S., et al. 2008, A\&A, 482, 585
\bibitem[Deharveng et al.(2010)]{Deharveng+10} Deharveng, L., Schuller, F., Anderson, L. D., et al. 2010, A\&A, 523, A6
\bibitem[Elmegreen \& Lada(1977)]{ElmegreenLada77}Elmegreen, B. G., \& Lada, C. J. 1977, ApJ, 214, 725
\bibitem[Ezawa et al.(2004)]{Ezawa+04} Ezawa, H., Kawabe, R., Kohno, K., \& Yamamoto, S. 2004, Proc. SPIE,
5489, 763
\bibitem[Ezawa et al.(2008)]{Ezawa+08} Ezawa, H., Kohno, K., Kawabe, R., et al. 2008, Proc. SPIE, 7012, 701208
\bibitem[Fukui et al.(2014)]{Fukui+14}  Fukui, Y., Ohama, A., Hanaoka, N., et al. 2014, ApJ, 780, 36
\bibitem[Fukui et al.(2015)]{Fukui+15} Fukui, Y., Harada, R., Tokuda, K., et al. 2015, ApJL, 807, L4
\bibitem[Fukui et al.(2016)]{Fukui+16}  Fukui, Y., Torii, K., Ohama, A., et al. 2016, ApJ, 820, 26
\bibitem[Fukui et al.(2017)]{Fukui+17} Fukui, Y., Torii, K., Hattori, Y., et a. al. 2017a, arXiv:1701.04669
\bibitem[Furukawa et al.(2009)]{Furukawa+09} Furukawa, N., Dawson, J. R., Ohama, A., et al. 2009, ApJL, 696, L115
\bibitem[Frerking et al.(1982)]{Frerking+82}Frerking, M. A., Langer, W. D., \& Wilson, R. W. 1982, ApJ, 262, 590
 
\bibitem[Goldreich \& Kwan(1974)]{GoldreichKwan74} Goldreich, P., \& Kwan, J. 1974, ApJ, 189, 441
\bibitem[Habe \& Ohta(1992)]{HabeOhta92} Habe, A., \& Ohta, K. 1992, PASJ, 44, 203
\bibitem[Haworth et al.(2015a)]{Haworth+15a} Haworth, T. J., Tasker, E. J., Fukui, Y., et al. 2015a, MNRAS,
450, 10
\bibitem[Haworth et al.(2015b)]{Haworth+15b} Haworth, T. J., Shima, K., Tasker, E. J., et al. 2015b, MNRAS,
454, 1634
\bibitem[Hosokawa \& Inutsuka(2006)]{HosokawaInutsuka06} Hosokawa, T., \& Inutsuka, S.-i. 2006, ApJ, 646, 240
\bibitem[Hydari-Malayeri(1988)]{Hydari-Malayeri88} Hydari-Malayeri, M. 1988, A\&A, 202, 240
\bibitem[Inoue \& Fukui(2013)]{InoueFukui13} Inoue, T., \& Fukui, Y. 2013, ApJL, 774, L31
\bibitem[Inoue et al.(2017)]{Inoue+17} Inoue, T., Hennebelle, P., Fukui, Y., et a. al. 2017, arXiv:1707.02035
\bibitem[Krumholz \& McKee(2008)]{KrumholzMcKee08} Krumholz, M. R., \& McKee, C. F. 2008, Nature, 451, 1082
\bibitem[Leung et al.(1984)]{Leung+84} Leung, C. M., Herbst, E., \& Huebner, W. F. 1984, ApJS, 56, 231
\bibitem[McKee \& Tan(2003)]{McKeeTan03} McKee, C. F., \& Tan, J. C. 2003, ApJ, 585, 850
\bibitem[Nakamura et al.(2012)]{Nakamura+12} Nakamura, F., Miura, T., Kitamura, Y., et al. 2012, ApJ, 746, 25
\bibitem[Nishimura et al.(2015)]{Nishimura+15} Nishimura, A., Tokuda, K., Kimura, K., et al. 2015, ApJS, 216, 18
\bibitem[Ohama et al.(2010)]{Ohama+10} Ohama, A., Dawson, J. R., Furukawa, N., et al. 2010, ApJ, 709, 975
\bibitem[Okamoto et al.(2017)]{Okamoto+17} Okamoto, R., Yamamoto, H., Tachihara, K., et al. 2017, ApJ, 838, 132
\bibitem[Pagani et al.(1993)]{Pagani+93} Pagani L., Heydari-Malayeri M., Castets A., 1993, A\&A 275, 573
\bibitem[Pe{\~n}aloza et al.(2017)]{Penaloza+17} Pe{\~n}aloza, C. H., Clark, P. C., Glover, S. C. O., et al. 2017, MNRAS,
465, 2277
\bibitem[Saigo et al.(2017)]{Saigo+17} Saigo, K., Onishi, T., Nayak, O., et al. 2017, ApJ, 835, 108 
\bibitem[Sano et al.(2017)]{Sano+17} Sano, H., Enokiya, R., Hayashi, K., et al. 2017, arXiv:1706.05763
\bibitem[Takahira et al.(2014)]{Takahira+14} Takahira, K., Tasker, E. J.,\& Habe, A. 2014, ApJ, 792, 63
\bibitem[Tan et al.(2014)]{Tan+14} Tan, J. C., Beltr{\'a}n, M. T., Caselli, P., et al. 2014, Protostars and
Planets VI, 149
\bibitem[Torii et al.(2015)]{Torii+15} Torii, K., Hasegawa, K., Hattori, Y., et al. 2015, ApJ, 806, 7
\bibitem[Torii et al.(2017)]{Torii+17} Torii, K., Hattori, Y., Hasegawa, K. et al. 2017, ApJ, 835, 142
\bibitem[van der Walt et al.(2012)]{vanderWalt+12} van der Walt D. J., de Villiers, H. M., \& Czanik, R. J. 2012, ApJ, 144, 13
\bibitem[Whitney et al.(2003)]{Whitney+03} Whitney, B. A., Wood, K., Bjorkman, J. E., \& Cohen, M. 2003, ApJ, 598, 1079 
\bibitem[Whitworth et al.(1994)]{Whitworth+94} Whitworth, A. P., Bhattal, A. S., Chapman, S. J., Disney, M. J., \& Turner, J. A. 1994, A\&A, 290, 421
\bibitem[Wilson \& Rood(1994)]{WilsonRood94} Wilson, T. L., \& Rood, R. 1994, ARA\&A, 32, 191
\bibitem[Yamaguchi et a.(1999)]{Yamaguchi+99} Yamaguchi, N., Mizuno, N., Saito., et al. 1999, PASJ, 51, 775
\bibitem[Yoda et al.(2010)]{Yoda+10} Yoda, T., Handa, T., Kohno, K., et al. 2010, PASJ, 62, 1277
\bibitem[Zavagno et al.(2007)]{Zavagno+07} Zavagno, A., Poma{\'e}s, M., Deharveng, L., et al. 2007, A\&A, 472, 835
\bibitem[Zavagno et al.(2010)]{Zavagno+10} Zavagno, A., Russeil, D., Motte, F., et al. 2010, A\&A, 518, L81
\bibitem[Zinnecker \& Yorke(2007)]{ZinneckerYorke07} Zinnecker, H., \& Yorke, H. W. 2007, ARA\&A, 45, 481
\end{thebibliography}
\end{document}